\documentclass[floats,floatfix,showpacs,amssymb,prd,twocolumn,superscriptaddress,nofootinbib]{revtex4-1}
\usepackage{amssymb,amsmath,verbatim,color,mathtools,needspace,enumitem,etoolbox,xcolor}
\usepackage{hyperref}
\definecolor{linkcolor}{rgb}{0.0,0.3,0.5}
\definecolor{urlcolor}{rgb}{0.27,0.55,0.}
\definecolor{funcolor}{rgb}{0.65, 0.16, 0.16}
\hypersetup{colorlinks=true,linkcolor=linkcolor,citecolor=linkcolor,filecolor=linkcolor,urlcolor=linkcolor}
\newcommand{\flag}[1]{\textsc{\lowercase{#1}}}
\newcommand{\fun}[1]{\textcolor{funcolor}{\texttt{#1}}}
\newcommand{\comp}[1]{\texttt{#1}}
\DeclareMathOperator\sign{sgn}
\newcommand{\precession}{\mbox{\flag{precession}}}
\newcommand{\python}{\flag{Python}}
\usepackage{etoolbox}
\makeatletter
\preto{\@verbatim}{\topsep=3.5pt \partopsep=4pt }
\makeatother
\usepackage{graphicx}
\usepackage{capt-of}

\def\araa{\rm{ARA\&A}}
\def\apj{\rm{ApJ}}
\def\apjl{\rm{ApJ}}
\def\apjs{\rm{ApJS}}

\def\mnras{\rm{MNRAS}}

\def\prd{\rm{PRD}}

\def\prl{\rm{PRL}}
\def\pasa{\rm{PASA}}

\def\ssr{\rm{Space~Sci.~Rev.}}

\def\nat{\rm{Nature}}

\def\physrep{\rm{Phys.~Rep.}}

\def\cqg{\rm{CQG}}

\begin{document}

\title{{\sc precession}: Dynamics of spinning black-hole binaries with python}
\author{Davide Gerosa}
\email{d.gerosa@damtp.cam.ac.uk}
\affiliation{Department of Applied Mathematics and Theoretical Physics, Centre for Mathematical Sciences, University of Cambridge, Wilberforce Road, Cambridge CB3 0WA, UK}

\author{Michael Kesden}
\email{kesden@utdallas.edu}
\affiliation{Department of Physics, The University of Texas at Dallas, Richardson, TX 75080, USA }


\date{\today}

\begin{abstract}

We present the numerical code \precession, a new open-source \python\ module to study the dynamics of precessing black-hole binaries in the post-Newtonian regime.  The code provides a comprehensive toolbox to (i) study the evolution of the black-hole spins  along their precession cycles, (ii) perform gravitational-wave-driven binary inspirals using both orbit-averaged and precession-averaged integrations, and (iii) predict the properties of the merger remnant through fitting formulas obtained from numerical-relativity simulations.  \precession\ is a ready-to-use tool to  add  the black-hole spin dynamics to larger-scale numerical studies such as  gravitational-wave parameter estimation codes, population synthesis models to predict gravitational-wave event rates, galaxy merger trees and  cosmological simulations of structure formation. \precession\ provides fast and reliable integration methods to propagate statistical samples of black-hole binaries from/to large separations where they form to/from small separations where they  become detectable, thus linking gravitational-wave observations of spinning black-hole binaries to their astrophysical formation history. The code is also a useful tool to compute initial parameters for numerical-relativity simulations targeting specific precessing systems. \precession\ can be installed from the \python\ Package Index, and it is freely distributed  under version control on \textsc{github}, where further  documentation is  provided.

\end{abstract}

\maketitle

\section{Introduction}
\label{intro}

Spinning black-hole (BH) binaries are remarkably interesting physical systems lying at the edge of fundamental physics and astronomy. 
Astrophysical BHs  are described by the Kerr \cite{1963PhRvL..11..237K}  solution of General Relativity  and are fully characterized by their mass and angular momentum, or spin. In a binary system, couplings between the BH spins and the binary's orbital angular momentum introduce secular dynamical features on top of the binary's orbital motion: the two spins and the orbital plane precess about the direction of the total angular momentum of the system \citep{1994PhRvD..49.6274A,1995PhRvD..52..821K}. Meanwhile, energy and  momentum are slowly dissipated away in the form of gravitational waves (GWs) and the orbital separation consequently shrinks  \citep{1963PhRv..131..435P}. GW-driven inspiral may ultimately lead to the merger of the two BHs.

The three phenomena highlighted above (orbit, precession and inspiral) take place on different timescales. While the two BHs orbit about each other with period $t_{\rm orb}\sim (r/r_g)^{3/2}$, the spins and the orbital angular momentum precess at the  rate $t_{\rm pre}\sim (r/r_g)^{5/2}$ and GW radiation reaction only affects the dynamics on  times $t_{\rm RR}\sim (r/r_g)^{4}$ (here $r$ is the binary separation and $r_g=GM/c^2$ is the gravitational radius of the total mass of the binary $M$). At separations $r\gg r_g$, the dynamics can be studied  successfully using the  post-Newtonian (PN) approximation to General Relativity (e.g. \cite{2014grav.book.....P}) and the three timescales are widely separated: $t_{\rm orb} \ll t_{\rm pre} \ll t_{RR}$. 
Multitimescale analyses can be used in this regime to efficiently disentangle the various dynamical features \citep{1995PhRvD..52..821K,2015PhRvL.114h1103K,2015PhRvD..92f4016G}. The timescale hierarchy breaks down, together with the entire PN approximation, at separations $r\sim r_g$ where the binary evolution can  be followed faithfully only using numerical-relativity simulations (see e.g.~\cite{2007arXiv0710.1338P}).

Spinning BHs now occupy a firm place in our understanding of the Universe.  
Astrophysical objects related to very energetic phenomena started being interpreted as BHs in the '60s \citep{1964ApJ...140..796S,1964SPhD....9..195Z} following the identification of the first quasar \citep{1963Natur.197.1040S} and the discovery of the first X-ray binary \citep{1972Natur.235...37W,1972Natur.235..271B}. BHs  are observed in two separated mass regimes: stellar-mass BHs, which are the endpoints of the life of some massive stars \citep{2014SSRv..183..223C}, and supermassive BHs, which reside at the center of most galaxies and help regulate their evolution \citep{1995ARA&A..33..581K}. Although challenging, robust spin measurements from electromagnetic observations are now possible in both mass regimes \citep{2013CQGra..30x4004R,2015PhR...548....1M}. 

BHs have been long predicted  to form binary systems: stellar-mass BH binaries are expected to form \emph{in the field} from the evolution of massive binary stars \citep{2014LRR....17....3P} and dynamically in dense stellar clusters  \citep{2013LRR....16....4B}; supermassive BH binaries are a natural by-product of hierarchical structure formation  and galaxy mergers \citep{1978MNRAS.183..341W,1980Natur.287..307B}. 
BH binaries are now an observational reality. Following challenging electromagnetic observations (see e.g. \cite{2006ApJ...646...49R} for a convincing candidate), the spectacular detection of GW150914 \citep{2016PhRvL.116f1102A} from the LIGO interferometers \citep{2016PhRvL.116m1103A} now constitutes  irrefutable astrophysical evidence of a merging stellar-mass BH binary. Merging supermassive BH binaries are the main targets of the future space-based GW interferometer eLISA \citep{2013arXiv1305.5720C,2016PhRvD..93b4003K} and current Pulsar Timing Arrays \citep{2009arXiv0909.1058J,2013PASA...30...17M,2013CQGra..30v4009K,2013CQGra..30v4010M}.

Spin precession is a crucial ingredient to both BH physics and GW astronomy. Although precessional modulations in the emitted GW signal require development of more elaborate waveforms \citep{2014PhRvD..89h4006P,2014PhRvL.113o1101H,2014PhRvD..89j4023C}, they constitute a promising channel to extract astrophysical information from GW observations \cite{2013PhRvD..87j4028G,2015ApJ...798L..17C,2016ApJ...818L..22A}.
Moreover, PN spin precession introduces complex dynamics to the final stage of BH inspirals \citep{2015PhRvL.114n1101L,2016PhRvD..93d4031L} and greatly affects the  properties of the BH remnants following binary mergers \citep{2010PhRvD..81h4054K,2010ApJ...715.1006K}.

In this paper, we present the numerical code \precession: an open-source \python\ module to study spinning BH binaries in the PN regime. In a nutshell, \precession\  performs BH binary inspirals tracking their precessional dynamics using both standard orbit-averaged  and new precession-averaged approaches. It also conveniently implements fitting formulas obtained from numerical-relativity simulations to predict
mass, spin and recoil of BH remnants following binary mergers.
\precession\ combines the flexibility of the high-level programming language \python\ with existing scientific libraries written in \textsc{C} and \textsc{Fortran} to bypass speed bottlenecks. 

Our code finds application in a variety of astrophysical problems.
Population synthesis models to predict GW rates (e.g.~\cite{2008ApJS..174..223B}) still lack the PN evolution of the BH spins which has been shown to critically depend on the  binary formation channel \citep{2013PhRvD..87j4028G}. Galaxy  merger trees (e.g.~\cite{2003ApJ...582..559V,2012MNRAS.423.2533B}) and  large-scale cosmological simulations (e.g.~\cite{2005Natur.435..629S,2014Natur.509..177V}) do not typically evolve the spin directions in the PN regime, although these are critical to address, e.g., the galaxy/BH occupation fraction \citep{2008MNRAS.384.1387V,2015MNRAS.446...38G} and the detectability of recoiling BHs \citep{2012AdAst2012E..14K,2016MNRAS.456..961B}.  We provide PN integrators to extend existing treatments of the astrophysical evolution of the BH spins \citep{2008ApJ...682..474B, 2010ApJ...719L..79F,2008ApJ...684..822B,2014ApJ...794..104S} through the GW-driven regime of the binary inspiral. The methods implemented in \precession\ to analyze the BH spin dynamics could provide initial parameters to numerical-relativity simulations (e.g. \citep{2009PhRvD..79b4003S,2015PhRvD..92j4028O}) targeting specific precessing systems. GW parameter-estimation codes (e.g. \citep{2015PhRvD..91d2003V}) may also benefit from our formulation of the spin-precession problem in terms of timescale separations. \precession\  can easily propagate BH binaries backwards from GW observation to arbitrarily large separation, thus reconstructing their entire inspiral history. Our multitimescale formulation of the problem could also help in the ongoing effort of building efficient GW templates for precessing systems \cite{2016arXiv160603117C}. Overall, we believe that \precession\ will be a useful tool to interpret numerical results and GW observations of precessing BH binaries and facilitate more accurate  modeling of their astrophysical environments.

This paper is organized as follows. Sec.~\ref{codeov} provides a general overview of the code; Sec.~\ref{spinpresec} is devoted to the spin precession dynamics; Sec.~\ref{GWinsp} describes the integration of the PN equations of motion to perform BH inspirals; Sec.~\ref{BHremnant} summarizes the implementation of numerical-relativity fitting formulas to predict the properties of postmerger BHs; Sec.~\ref{examples} contains various practical examples to use \precession;  Sec.~\ref{concl} highlights our conclusions and anticipates future features of the code. From now on, equations are written in geometrical  units $(c=G=1)$. As specified in Sec.~\ref{installation}, code units also set the binary's total mass to 1.

\section{Code Overview}
\label{codeov}

In this section we give a general overview of the code. Sec.~\ref{installation} describes code installation;  Sec.~\ref{minimal} presents a minimal working example; Sec.~\ref{secdocumentation} provides details on documentation and source distribution; Sec.~\ref{units} describes units and parallel programming features.

\subsection{Installation}
\label{installation}

 \precession\ is a {\sc python}  \citep{Python} module and is part of the {\sc python} Package Index:   \href{https://pypi.python.org/pypi/precession}{pypi.python.org/pypi/precession}. The code can be installed in a single line through the  package management system \comp{pip}:
\begin{verbatim}
    pip install precession
\end{verbatim}
Useful options to the command above include \comp{--user} for users without root privileges and \comp{--upgrade} to update a preexisting installation. The scientific libraries \textsc{numpy} \citep{Walt}, \textsc{scipy} \citep{Jones:2001aa}, \textsc{matplotlib} \citep{2007CSE.....9...90H} and \textsc{parmap} \citep{Oller} are specified as prerequisites and, if not present, will be  installed/updated  together with \precession. 
 \precession\ has been tested on  \python\ 2.7 distributions; porting to \python\ 3 is under development. 
 
 Once \precession\ has been installed, it has to be imported typing
\begin{verbatim}
    import precession
\end{verbatim}
from within a \python\ console or script. The main module \fun{precession} contains $\sim 80$ functions for a total of $\sim 1700$ code lines. The submodule \fun{precession.test} consists of $\sim 300$ code lines divided into seven examples routines. If needed, this has to be imported separately typing
\begin{verbatim}
    import precession.test
\end{verbatim}
All functions and examples that should be called by the user are described in this paper.

\subsection{A first working example}
\label{minimal}

\begin{figure}
\includegraphics[width=\columnwidth]{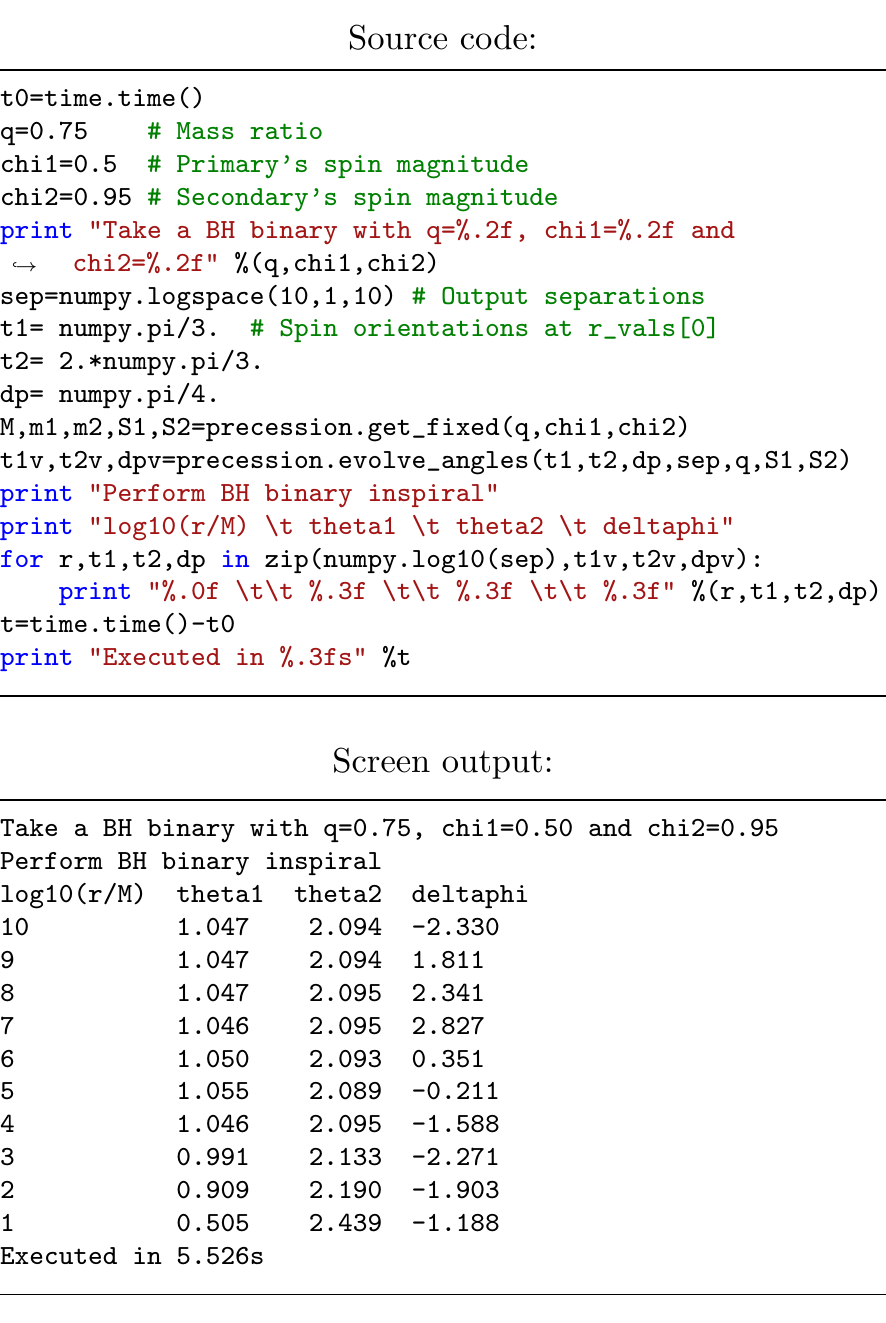}
\caption{Source code (top) and screen output (bottom) of the example \fun{test.minimal} described in Sec.~\ref{minimal}.  We select a BH binary at \mbox{$r=10^{10}M$} and track the directions of the two spins and the orbital angular momentum [cf. Eqs.(\ref{angdef1})-(\ref{signdp})] during its PN inspiral till \mbox{$r=10M$}. We use precession-averaged PN equations, which require random samplings of the precessional phase, see Sec.~\ref{GWinsp} (different code executions will therefore return different values of the spin angles). The execution time reported is obtained  using a single core of a 2013 Intel i5-3470 3.20GHz CPU. These lines can be executed typing \comp{precession.test.minimal()}. \label{minimalcode}}
\end{figure}

A minimal working example of some features of \precession\ is shown in Fig.~\ref{minimalcode}. We encourage the reader to execute this code snippet typing
\begin{verbatim}
    precession.test.minimal()
\end{verbatim}
We initialize a BH binary at the extremely large separation of 10 billion gravitational radii ($r=10^{10} M$) and  evolve it down to small separations  ($r=10M$) where the PN approximation breaks down. The integration is performed using precession-averaged PN equations of motion, as described later in Sec.~\ref{precav}. The evolution of the BH spins along such an enormous separation range is computed in less than six seconds using a single core of a standard off-the-shelf desktop machine.

\subsection{Documentation and source distribution}
\label{secdocumentation}
This paper describes the numerical code \precession\ in its v1.0 release. The code is under active development and additional features will be added regularly. Earlier versions of the code were used in the following published results: \cite{2015PhRvL.115n1102G,2015PhRvD..92f4016G,2015MNRAS.451.3941G,2015PhRvL.114h1103K,2015MNRAS.446...38G,2016PhRvD..93d4071T,2016arXiv160604226G}.

The source code is distributed under \textsc{git} version-control system at 
\vspace{-0.1cm}
\begin{quote}
\hspace{-0.4cm}\href{https://github.com/dgerosa/precession}{github.com/dgerosa/precession}$\;\;$(code),
\end{quote}
\vspace{-0.1cm}
and it is released under the \href{https://creativecommons.org/licenses/by/4.0/}{CC BY 4.0} license. Extensive code documentation can be generated automatically in html format from \python's docstrings using the text processor \textsc{pdoc} \citep{pdoc}. Documentation is regularly uploaded to a dedicated branch of the \textsc{git} repository and it is available at 
\vspace{-0.1cm}
\begin{quote}
\hspace{-0.4cm}\href{https://dgerosa.github.io/precession}{dgerosa.github.io/precession}$\;\;$(documentation).
\end{quote}
\vspace{-0.1cm}
The same information can also be accessed using {\python}'s built-in help system, e.g. \comp{help(precession.function)}.  
Additional resources and results are available at \href{http://davidegerosa.com/precession}{davidegerosa.com/precession}.

\subsection{Units and parallel features}
\label{units}
All quantities in  the code must be specified in total-mass units, i.e. $c=G=M=1$. For instance, the code variable for the binary separation \comp{r} stands for $r c^2/GM$; equivalently, the angular-momentum magnitude variable \comp{L} stands for $cL/GM^2$.

\precession\ includes some parallel programming features. Embarrassingly parallel  tasks, such as computing several PN inspirals (Sec.~\ref{GWinsp}), are sent to different cores to speed up the computation. By default, \precession\ autodetects the number of available cores in the executing machine and splits the operations accordingly. 
Parallel execution can be controlled using the global integer variable \fun{CPUs}, which specifies the number of parallel processes. For instance, serial execution can be enforced setting \comp{CPUs=1} (cf. Sec.~\ref{timing}).

Outputs of some functions are automatically stored, such that further executions of code scripts do not require full recalculation. The location of the output directory is controlled by the global string variable \fun{storedir}, which is set by default to \comp{"./precession\_checkpoints"}.  The output directory is automatically created if needed, or can be created manually using \fun{make\_temp}. Stored data files can be deleted using \fun{empty\_temp}.

\section{Spin precession}
\label{spinpresec}

In this section we present how to use \precession\ to study BH binaries on the spin precession timescale  where GW emission can be neglected. After introducing double-spinning BH binaries (Sec.~\ref{mainpar}), we describe two useful parametrizations of the precession dynamics (Sec.~\ref{params}) and  discuss their constraints (Sec.~\ref{params}). Time evolution of BH binaries along their precession cycles is described in Sec.~\ref{timeevprec}. Finally, Sec.~\ref{morph} shows how to classify  BH binaries according to their precessional morphologies.

\subsection{Black-hole binaries in the post-Newtonian regime}
\label{mainpar}

Throughout this paper we only consider  BH binaries on quasi-circular orbits. Astrophysical BH binaries are expected to circularize at large separation \citep{1963PhRv..131..435P, 1964PhRv..136.1224P} and the first GW detection confirms this finding \citep{2016arXiv160203840T}. However, eccentricity may be relevant for stellar-mass BH binaries formed in globular clusters \citep{2013LRR....16....4B} and supermassive BH binaries interacting with dense stellar environments \citep{2010ApJ...719..851S,2012JPhCS.363a2035R}. 
Generalization to eccentric orbits is an important extension of \precession, which is left to future work.

We use standard notation where the component masses $m_1$ and $m_2$ are combined into total mass $M=m_1+m_2$, mass ratio $q=m_2/m_1 \leq 1$ and symmetric mass ratio $\eta=m_1 m_2/ M^2 =q/(1+q)^2$; the spin magnitudes $S_i=m_i^2 \chi_i$  (hereafter $i=1,2$) are given in terms of the dimensionless spin parameters $0\leq \chi_i \leq 1$. The magnitude of the orbital angular momentum $\mathbf{L}$ is related to  the binary separation $r$ through the Newtonian expression $L= m_1 m_2 \sqrt{r/M}$. 
The utility \fun{get\_fixed} provides the component masses $m_i$ and the spin magnitudes $S_i$ in terms of  $q$ and $\chi_i$ in code units; similarly, \fun{get\_L} returns the Newtonian expression for the magnitude of the orbital angular momentum.

Before proceeding with the code implementation, we point out that  \precession\ is explicitly designed to handle genuine double-spin physics. Nonspinning and single-spin binaries (i.e. $\chi_1=0$ and/or $\chi_2=0$) represent singular cases that cannot be handled with the present version of the code.
In practice, these systems can be well approximated by setting $\chi_i\gtrsim 0.001$. 

\precession\  loses accuracy in the extreme-mass-ratio limit $q\to 0$ (where other methods are required to study the dynamics, e.g. \cite{2011LRR....14....7P}) and the equal-mass limit $q\to 1$ [where the parametrization chosen to describe the precession cycle breaks down, e.g. Eq.~(\ref{tran2})]. Our results have been well tested in the regime $0.005\lesssim q \lesssim 0.995$. \precession\ currently features an alternative implementation to study the strictly equal-mass case $q=1$, which exploits additional constants of motion  \citep{2008PhRvD..78d4021R,2014PhRvD..89j4052L}.  These findings will be presented elsewhere  \cite{VosmeraPrep}.
\subsection{Parametrization of double spin precession}
\label{params}

The time evolution of the three vectors $\mathbf{S_1}$, $\mathbf{S_2}$ and $\mathbf{L}$ in an inertial frame is a nine-parameter problem. However, only four parameters are needed to describe the relative orientations of the three momenta \citep{2004PhRvD..70l4020S,2006PhRvD..74j4005B,2014PhRvD..89l4025G}.
One of these parameters is the orbital separation $r$ (or equivalently the magnitude $L$), which is constant on $t_{\rm pre}$ and decreases on $t_{\rm RR}$ because of GW emission. Two possible choices for the remaining three degrees of freedom are:

\begin{enumerate}
\item 
The spin directions can be described in terms of three angles,
\begin{align}
\label{angdef1}
\cos\theta_1 &= \mathbf{\hat S_1} \cdot \mathbf{\hat L} \,,
\\
\label{angdef2}
 \cos\theta_2&=\mathbf{\hat S_2} \cdot \mathbf{\hat L}\,,
\\
\cos\Delta\Phi&=\frac{\mathbf{\hat S_1} \times \mathbf{\hat L}}{|\mathbf{\hat S_1} \times  \mathbf{\hat L} |} \cdot 
\frac{\mathbf{\hat S_2} \times \mathbf{\hat L}}{|\mathbf{\hat S_2} \times \mathbf{\hat L} |},
\label{angdef3}
\end{align}
where the sign of $\Delta\Phi$ is chosen such that
\begin{align}\label{signdp}
\sign\Delta\Phi = \sign\{ \mathbf{L} \cdot [(\mathbf{S_1} \times \mathbf{L}) \times (\mathbf{S_2} \times \mathbf{L})] \}.
\end{align}
In other words, $\theta_1$ and $\theta_2$ are the angles between the two spins and the orbital angular momentum (\emph{tilt angles}) and $\Delta\Phi$ is the angle between the projections of the two spins onto the orbital plane (see Fig.~1 in \cite{2014PhRvD..89l4025G}). Despite being very intuitive, this description makes the understanding of the underlying phenomenology rather complicated because all three variables $(\theta_1,\theta_2,\Delta\Phi)$ vary on both the precession and the inspiral timescales.

\item A more physical choice can be made to exploit  the timescale separation $t_{\rm pre}\ll t_{RR}$. The magnitude of the total angular momentum 
\begin{align}
J=|\mathbf{L}+\mathbf{S_1}+\mathbf{S_2}| 
\label{defJ}
\end{align}
is conserved on the timescale $t_{\rm pre}$ where GW emission can be neglected. Moreover, the projected effective spin \citep{2001PhRvD..64l4013D,2008PhRvD..78d4021R}
\begin{equation}\xi \equiv M^{-2} [(1+q)\mathbf{S}_1 + (1+q^{-1})\mathbf{S}_2] \cdot \hat{\mathbf{L}}
\label{defxi}
\end{equation}
is a constant of motion of  the (orbit-averaged) 2PN spin-precession and 2.5PN radiation-reaction equations (cf. Sec.~\ref{orbav}) and is therefore conserved on both $t_{\rm pre}$ and $t_{\rm RR}$.
This implies that the entire dynamics  on $t_{\rm pre}$ can be encoded in a single variable, which can be chosen\footnote{Equivalently, one can choose  the angle $\varphi'$ defined in Eq.~(9) of \cite{2015PhRvD..92f4016G}. \precession\ contains additional routines to analyze the dynamics in terms of this angle. The most relevant functions are called  \fun{get\_varphi} and \fun{region\_selection}.} to be  the magnitude of the total spin \citep{2015PhRvL.114h1103K}
\begin{align}
S=|\mathbf{S_1}+\mathbf{S_2}|\,.
\label{defS}
\end{align}
\end{enumerate}
The two descriptions --in terms of $(\theta_1,\theta_2,\Delta\Phi)$ and $(\xi,J,S)$-- are related by the following sets of transformations 
\begin{align}
&\begin{dcases}
S = [S_1^2 + S_2^2 + 2S_1S_2(\sin\theta_1\sin\theta_2\cos\Delta\Phi \\
\qquad\qquad\qquad+ \cos\theta_1\cos\theta_2)]^{1/2}~, \\
J = [L^2 +  S^2 + 2L (S_1 \cos\theta_1 + S_2 \cos\theta_2)]^{1/2}~,\\
\xi = \frac{1+q}{q M^2}(qS_1\cos\theta_1 + S_2\cos\theta_2)~;
\end{dcases}
\label{tran1}\\
&\begin{dcases}
\cos\theta_1 =  \frac{1}{2(1-q)S_1} \left[ \frac{J^2 - L^2 -S^2}{L} - \frac{2qM^2\xi}{1+q} \right]\,, \\
\cos\theta_2 =  \frac{q}{2(1-q)S_2} \left[ -\frac{J^2 - L^2 -S^2}{L} + \frac{2M^2\xi}{1+q} \right]\,, \\
\cos\Delta\Phi = \frac{1}{\sin\theta_1\sin\theta_2} \left({\frac{S^2 - S_1^2 - S_2^2}{2S_1S_2} - \cos\theta_1\cos\theta_2}\right),
\end{dcases}
\label{tran2}
\end{align}
which are implemented in \fun{from\_the\_angles} and \fun{parametric\_angles}. Similarly, Eqs.~(\ref{angdef1})-(\ref{signdp}) can be evaluated using \fun{build\_angles}.
The angle  $\theta_{12}=\arccos {\mathbf{ \hat S_1}}  \cdot {\mathbf{\hat S_2}}$  between the two spins can be computed using both sets of variables:
\begin{align} 
\cos\theta_{12} &= \frac{S^2 - S_1^2 - S_2^2}{2S_1S_2}
\notag
\\
 &= \sin\theta_1\sin\theta_2\cos\Delta\Phi + \cos\theta_1\cos\theta_2\,.
\end{align}
Equations (\ref{tran1}) and (\ref{tran2}) do not depend on the sign of $\Delta\Phi$. This reflects the symmetry of the dynamics between the first and second half of the precession cycle (cf.~Sec.~\ref{timeevprec}).
If the spin vectors are available in the current computation (e.g. from orbit-averaged evolutions, see Sec.~\ref{orbav}), \precession\
evaluates the sign of $\Delta\Phi$ directly from Eq.~(\ref{signdp}). If this is not the case, $\sign\Delta\Phi$ must be specified by the user according to the evolution of $S$, as in the example of Sec.~\ref{spansec}. In case of precession-averaged inspirals (Sec.~\ref{phaseresamp}), the sign of $\Delta\Phi$ is assigned randomly.

\subsection{Geometrical constraints}
\label{geocon}

The physical range of the three angles $(\theta_1,\theta_2,\Delta\Phi)$ is given by the independent constraints $\theta_1\in[0,\pi]$, $\theta_1\in[0,\pi]$ and $\Delta\Phi\in[-\pi,\pi]$. Geometrical constraints on $\xi$, $J$, $S$ can be derived from Eqs.~(\ref{defxi})-(\ref{defS}) and read
 \begin{gather}
- (1+q) (   {S}_1 +  {S}_2/q) \,\leq \, M^2 \xi\, \leq \, (1+q) (   {S}_1 +  {S}_2/q)\,, \label{xilim}
\\
\max(0 , L - S_1 - S_2, |S_1 - S_2| - L)\,\leq \,J\, \leq  L+S_1+S_2\,, \label{Jlim}
\\
|S_1 - S_2|\,\leq \,S\, \leq  S_1+S_2\,.
\label{ssolim}
\end{gather}
Equations (\ref{xilim}), (\ref{Jlim}) and (\ref{ssolim}) are returned by \fun{xi\_lim}, \fun{J\_lim} and \fun{Sso\_limits}, respectively. These constraints are not independent of each other. For a given $J$ satisfying Eq.~(\ref{Jlim}), the magnitude $S=|\mathbf{S_1}+\mathbf{S_2}|=|\mathbf{J}-\mathbf{L}|$ has to satisfy 
\begin{equation}
\max(|J-L|, |S_1 - S_2|) \leq S \leq  \min(J+L, S_1 + S_2) \,,
\label{stlim}
\end{equation} 
which is given by \fun{St\_limits}. Allowed values of $\xi$ are then given by 
\begin{align}
\min_S \xi_-(S) \,\leq \,&\xi\, \leq \, \max_S \xi_+(S) \,,\label{xiallowed}
\end{align}
where $\xi_\pm$ are the \emph{effective potentials} for BH binary spin precession  \citep{2015PhRvL.114h1103K}
\begin{align} \label{effpot}
&\xi_\pm(S) = \{ (J^2 - L^2 - S^2)[S^2(1+q)^2 - (S_1^2 - S_2^2)(1- q^2)] \notag \\
&\quad  \pm (1- q^2) \sqrt{ [J^2 - (L - S)^2] [(L + S)^2 - J^2]}  \notag \\
&\quad  \times \sqrt{[S^2 - (S_1 - S_2)^2][(S_1 + S_2)^2 - S^2]}\}\big/(4qM^2S^2L)\,.
\end{align}
In Ref. \cite{2015PhRvD..92f4016G} we proved that  $\xi_{+}$ ($\xi_{-}$) admits a single maximum (minimum) within the range of $S$ given by Eq.~(\ref{stlim}) for any value of $J$ satisfying Eq.~(\ref{Jlim})\footnote{One can also prove that $\min_S \xi_-(S)=\max_S \xi_+(S)$ if and only if $J=L+S_1+S_2$ or $J=\max(0, L-S_1-S_2, |S_1-S_2|-L)$ \citep{2015PhRvD..92f4016G}. Only one value of $\xi$ is allowed in these peculiar cases and can be evaluated using \fun{xi\_at\_Jlim}.}. The extremization of the effective potentials 
 is performed in \fun{xi\_allowed} using  \comp{scipy.optimize.fminbound} with a bracketing interval given by Eq.~(\ref{stlim}). 
Analogously,  \fun{J\_allowed} computes the allowed range of $J$ for any value of  $\xi$ satisfying Eq.~(\ref{xilim}). 
If needed, the effective potentials of Eq.~(\ref{effpot}) can be evaluated directly using \fun{xi\_plus} and \fun{xi\_minus}; their derivatives $d \xi_\pm/ dS$   are implemented in  \fun{dxidS\_plus} and \fun{dxidS\_minus}. 

Once consistent values of $J$ and $\xi$ have been selected (cf. Sec.~\ref{parselsec} for a practical example), the binary dynamics on $t_{\rm pre}$ is fully encoded in the evolution of $S$. The magnitude $S$ oscillates between the two  solutions $S_{\pm}$ of the equations $\xi_\pm(S)=\xi$. A precession cycle therefore consists of a complete oscillation $S_- \to S_+\to S_-$. The radical equations \mbox{$\xi_\pm(S)=\xi$} are solved in \fun{Sb\_limits} using 
\comp{scipy.optimize.brentq}. From  experiments in wide regions of the parameter space, we report a numerical  accuracy of $\Delta S_\pm/M^2\sim  10^{-8}$.

The two roots $S_\pm$ coincide at the extrema of the effective potentials \mbox{$\xi=\min_S \xi_-(S)$} and \mbox{$\xi=\max_S \xi_+(S)$}, where consequently  the magnitude of the total spin $S$ remains constant. These are peculiar configurations where the \emph{relative} orientation of  $\mathbf{S_1}$, $\mathbf{S_2}$ and $\mathbf{L}$ does not evolve on $t_{\rm pre}$. It is straightforward to prove that they are characterized by $\sin\Delta\Phi=0$: the three angular momenta share the same plane and jointly precess about the direction of $\mathbf{J}$. These solutions have been discovered more than a decade ago by Schnittman \cite{2004PhRvD..70l4020S} and called  \emph{spin-orbit resonances} (for other studies see \cite{2014CQGra..31j5017G,2016MNRAS.457L..49C}).  One can prove that extremizing the effective potential $\xi_\pm$ is  equivalent to solving Eq.~(3.5) of \cite{2004PhRvD..70l4020S}.
Two spin-orbit resonances are present for any value of $\xi$: they are characterized by $\Delta\Phi=0$ and $\Delta\Phi=\pi$ and correspond to the largest and lowest values of $J$ compatible with the chosen $\xi$ (cf. Fig.~5 in \cite{2015PhRvD..92f4016G}). The angles $\theta_1$ and $\theta_2$ corresponding to both resonances  $\Delta\Phi=0,\pi$ can be evaluated using \fun{resonant\_finder}. 

The values of $J$ and $\xi$ corresponding to the four (anti)aligned configurations $\cos\theta_i=\pm 1$ are  returned by \fun{aligned\_configurations}. The thresholds of the precessional instability discovered in \cite{2015PhRvL.115n1102G} are returned by \fun{updown}. 

\subsection{Binary evolution on the precession timescale}
\label{timeevprec}

The rate of variation of $S$  between the two extrema $S_\pm$,
\begin{align} 
\frac{dS}{dt} &= -\frac{3(1-q^2)}{2q} \frac{S_1S_2}{S} \frac{(\eta^2M^3)^3}{L^5} \left( 1 - \frac{\eta M^2 \xi}{L}  \right) 
\notag \\
&\quad\quad\;\;\times \sin\theta_1\sin\theta_2\sin\Delta\Phi\,
 \label{dSdt}
\\
&=\pm\frac{3}{2}\eta M \left[ 1 - \xi\left(\frac{r}{M}\right)^{-1/2} \right] \left( \frac{r}{M} \right)^{-5/2}
\notag \\
&\quad\quad\;\;\times\sqrt{(\xi_+ - \xi)(\xi - \xi_-)}
\end{align}
 follows directly from the 2PN spin-precession equations [here reported in Eqs.~(\ref{dmomdt})-(\ref{precession2}), see \cite{1995PhRvD..52..821K}] and can be evaluated using \fun{dSdt}. The solutions $S_\pm$ of the equations $\xi_\pm(S)=\xi$ correspond to turning points in the evolution of $S$, i.e. $dS/dt=0$.
The time evolution of a BH binary during (half of) a precession cycle is  given by the integral
\begin{align}
t(S)= \int_{S_-}^{S} \frac{dS'}{|dS'/dt|}\,, \quad S\in [S_-, S_+]\,.
\label{tofS}
\end{align}
The integrand $|dt/dS|^{-1}$ is regular everywhere in $S\in (S_-, S_+)$, while the limits \begin{align}
\lim_{S \to S_\pm}\frac{1}{|dS/dt|} \propto \frac{1}{\sqrt{|S - S_\pm|}} \label{integrablesings}
\end{align}
ensure integrability\footnote{The only exception is the up-down configuration ($\cos\theta_1=1$, $\cos\theta_2=-1$) in its instability region, where $\tau \to \infty$ \citep{2015PhRvL.115n1102G}.} at $S_\pm$. The numerical integration of Eq.~(\ref{tofS}) is performed in \fun{t\_of\_S} and its inverse \fun{S\_of\_t}, using  standard quadrature through  \comp{scipy.integrate.quad}. Equation (\ref{tofS}) can used to reparametrize the binary dynamics in terms of  time (cf. Sec.~\ref{spansec}). The precessional period $\tau$ is defined as the time for a complete precession cycle $S_- \to S_+\to S_-$,
\begin{align}
\tau= 2 \int_{S_-}^{S^{+}} \frac{dS'}{|dS'/dt|}\,,
\label{tau}
\end{align}
and can be computed  using   \fun{precessional\_period}. 

The direction of $\mathbf{J}$ is  constant as long as radiation reaction is being neglected. The orbital angular momentum $\mathbf{L}$  precesses about that fixed direction at a rate \citep{2015PhRvL.114h1103K} 
\begin{align} \label{omegaz}
\Omega_z 
&=\frac{J}{2} \left( \frac{\eta^2M^3}{L^2} \right)^3 \bigg\{ 1 + \frac{3}{2\eta} \left( 1 - \frac{\eta M^2 \xi}{L} \right)
\notag \\ 
& -\frac{3(1+q)}{2q
} \left(1 - \frac{\eta M^2 \xi}{L} \right)[4(1-q)L^2(S_1^2 - S_2^2)
\notag \\
& -(1+q)(J^2 - L^2 -S^2)(J^2 - L^2 -S^2 - 4\eta M^2L\xi)]
\notag \\
&\times[J^2 - (L - S)^2]^{-1}[(L + S)^2 - J^2]^{-1} \bigg\}\,.
\end{align}
The vector  $\mathbf{L}$, therefore, spans an angle 
\begin{align}
\label{eqalpha}
\alpha = 2 \int_{S_-}^{S_+}  \Omega_z \frac{dS}{|dS/dt|}
\end{align}
about $\mathbf{J}$ during each precession cycle.
Equations~(\ref{omegaz}) and (\ref{eqalpha}) can be evaluated using \fun{Omegaz} and \fun{alphaz}, respectively. The azimuthal angle of the projection of $\mathbf{L}$ onto a plane orthogonal to $\mathbf{J}$ can be tracked using  \fun{alpha\_of\_S}, cf. Eq.~(30) of \cite{2015PhRvD..92f4016G}. The conditions $\alpha=2 \pi n$ ($n$ integer) correspond to  configurations where the precession frequency of $\mathbf{L}$ about $\mathbf{J}$ and that of the two spins are in resonance with each other \cite{ZhaoPrep}. Tools to analyze such peculiar configurations will be made available in future versions of the code.

\subsection{Spin morphologies}
\label{morph}
As discussed at great length in \cite{2015PhRvD..92f4016G}, the precessional behavior of spinning BH binaries can be classified in terms of three different {\it morphologies}. These are related to the evolution of $\Delta\Phi$ during a precession cycle. In particular, three situations are possible:
\begin{enumerate}
\item $\Delta\Phi$ circulates through the full range $[-\pi,+\pi]$;
\item $\Delta\Phi$ librates about $0$ (and never reaches $\pm\pi$);
\item $\Delta\Phi$ librates about $\pm\pi$ (and never reaches $0$).
\end{enumerate}
Examples of BH binaries in the different morphologies are studied in  Sec.~\ref{parselsec}. 
The spin-orbit resonances $\xi=\min_S(\xi_-)$ and $\xi=\max_S(\xi_+)$ can be interpreted as the limits of the two librating morphologies: as the precession amplitude $(S_+-S_-)$ goes to zero, $\Delta\Phi$ approaches one of the resonant configurations and locks onto either $0$ or $\pm \pi$ \citep{2004PhRvD..70l4020S}. 
The spin morphology is an interesting dynamical feature of BH binaries because, while it characterizes spin precession, it does not vary on the precession timescale (i.e., it is independent of $S$). 
Radiation reaction causes morphological transitions which are promising GW observables \citep{2014PhRvD..89l4025G,2016PhRvD..93d4071T}. Morphological classification is implemented in \fun{find\_morphology}. 

The loop formed by the two effective potentials $\xi_\pm$ of Eq.~(\ref{effpot}) encloses all binary configurations $(\xi,S)$ compatible with fixed values of $r$, $J$, $q$ and $S_i$. Regions of binaries with different morphologies can coexist in this plane in the following way (see Fig.~4 of \cite{2015PhRvD..92f4016G}):\begin{enumerate}
\item a single region where all  binaries librate about $\Delta\Phi=\pm\pi$;
\item two regions of binaries librating about $\Delta\Phi=\pm\pi$ separated by a third region of circulating binaries;
\item three different regions, where binaries librate about $\Delta\Phi=0$,  circulate  and  librate about $\Delta\Phi=\pm\pi$.
\end{enumerate}
This distinction is performed by \fun{phase\_xi}. A useful tool is provided in \fun{phase\_checker}, which ensures that the output of \fun{phase\_xi}   satisfies the constraints of Sec.~\ref{geocon}.

\section{Gravitational-wave-driven inspiral}
\label{GWinsp}

In this section we illustrate how to use \precession\ to compute BH inspirals. We provide a standard integrator of the orbit-averaged PN equations (Sec.~\ref{orbav}) and a framework to evolve binaries using our innovative precession-averaged approach (Sec.~\ref{precav}). A key ingredient is the  statistical resampling of the precessional phase, which is illustrated in Sec.~\ref{phaseresamp}. Finally, we present a new hybrid approach  where precessional cycles are tracked only during the last part of the inspiral (Sec.~\ref{hybrid}).

\vspace{0.1cm}
\subsection{Orbit-averaged evolutions}
\label{orbav}

GW emission dissipates energy and angular momentum, thus decreasing the binary separation. Following the seminal studies of Apostolatos {\it et al.} \cite{1994PhRvD..49.6274A} and Kidder \cite{1995PhRvD..52..821K}, the PN equations of motion for precessing systems have historically been studied averaging  over the orbital motion \citep{2004PhRvD..70l4020S,2008PhRvD..78d4021R,2009PhRvD..79j4023A,2011PhRvD..84d9901A,2006PhRvD..74j4033F,2006PhRvD..74j4034B}, which exploits the  inequalities $t_{\rm orb}\ll t_{\rm pre}$ and $t_{\rm orb}\ll t_{\rm RR}$. We provide a numerical integrator for the following set of orbit-averaged PN equations:

\begin{widetext}
\vspace{-0.65cm}
\begin{align}
\label{dmomdt}
\frac{d \mathbf{ S_1}}{dt} &= \mathbf{\Omega_1}\times \mathbf{ S_1}, 
\qquad
\frac{d \mathbf{ S_2}}{dt} = \mathbf{\Omega_2}\times \mathbf{ S_2}, 
\qquad 
\frac{d \mathbf{\hat L}}{dt} = -\frac{v}{\eta M^2}
\frac{d }{dt} ( \mathbf{ S_1}+ \mathbf{ S_2});
\\ \label{precession1}
M \mathbf{\Omega_1}&= \eta v^5 \left( 2 +\frac{3q}{2} \right) \mathbf{\hat L}
+  \frac{v^6}{2M^2}\left[  \mathbf{ S_2} - 3 \left(\mathbf{\hat L} \cdot \mathbf{ S_2} \right) \mathbf{\hat L}
- 3q \left(\mathbf{\hat L} \cdot \mathbf{ S_1} \right) \mathbf{\hat L}\right];
\\ \label{precession2}
\qquad 
M \mathbf{\Omega_2}&= \eta v^5 \left( 2 +\frac{3}{2q} \right) \mathbf{\hat L}
+  \frac{v^6}{2M^2}\left[\mathbf{ S_1} - 3 \left(\mathbf{\hat L} \cdot \mathbf{ S_1} \right) \mathbf{\hat L}
- \frac{3}{q} \left(\mathbf{\hat L} \cdot \mathbf{ S_2} \right) \mathbf{\hat L}\right];
\end{align}
\vspace{-0.35cm}
\begin{align}
\frac{dv}{dt}&= \frac{32}{5}\frac{\eta}{M}v^9 \Bigg\{ 1-
 v^2 \frac{743 +924 \eta }{336}+ 
 v^3 \Bigg[ 4 \pi - \sum_{i=1,2}
  \chi_i (\mathbf{\hat S_i} \cdot\mathbf{\hat L}   )\left(\frac{113}{12}\frac{m_i^2}{M^2} + \frac{25}{4}\eta\right)  \Bigg]
+  v^4
\Bigg[\frac{34103}{18144}+\frac{13661}{2016}\eta 
 +\frac{59}{18}\eta^2  
 \notag \\&
+ \frac{\eta \chi_1 \chi_2}{48} \left( 721(\mathbf{\hat S_1} \cdot\mathbf{\hat L})(\mathbf{\hat S_2} \cdot\mathbf{\hat L})
-247 (\mathbf{\hat S_1}\cdot\mathbf {\hat S_2} ) \right)
+ \frac{1}{96} \sum_{i=1,2}
\left(\frac{m_i \chi_i}{M}\right)^2 \left( 719 (\mathbf{\hat S_i} \cdot\mathbf{\hat L})^2 - 233 \right)
\Bigg]-
 v^5 \pi \frac{4159 +15876 \eta}{672} 
  \notag \\
&+ v^6\Bigg[\frac{16447322263}{139708800}+\frac{16}{3}\pi^2 -\frac{1712}{105}\left(\gamma_E +\ln4v\right)+ 
\left( \frac{451}{48} \pi^2 - \frac{56198689}{217728} \right) \eta
+\frac{541}{896}\eta^2 -\frac{5605}{2592}\eta^3\Bigg]
\notag \\& + v^7 \pi \Bigg[ -\frac{4415}{4032}+\frac{358675}{6048}\eta +\frac{91495}{1512}\eta^2\Bigg]
+O(v^8)\Bigg\} ; 
\label{radiationreaction}
\end{align}
\end{widetext}
where $v=\sqrt{M/r}$ is the orbital velocity and $\gamma_E\simeq0.577$ is Euler's constant. The spin-precession equations (\ref{dmomdt})-(\ref{precession2}) are accurate up to 2PN; corrections to the  radiation-reaction equation (\ref{radiationreaction}) are included up 3.5PN (2PN)  for \mbox{(non)spinning} terms
\citep{1963PhRv..131..435P,1964PhRv..136.1224P,1993PhRvD..47.4183K,1994PhRvD..49.6274A,
1995PhRvD..52..821K,1998PhRvD..57.5287P,2000PhRvD..61b4035G,2002PhRvD..65f1501B,2005PhRvD..71l9902B,2004PhRvL..93i1101B,2006PhRvD..74j4034B,2006PhRvD..74j4033F,2008PhRvD..78d4021R}.
Higher-order PN corrections to spin precession \citep{2013CQGra..30e5007M,2013CQGra..30g5017B,2015CQGra..32s5010B} and radiation reaction \citep{2014PhRvD..89f4058D,2016PhRvD..93h4037B} are not implemented in the current version of \precession\ (see also \cite{2016PhR...633....1P})
The importance of such additional corrections on the conservation of $\xi$ and their quantitative effect at small separations is still unclear and surely merits further investigation.

Orbit-averaged inspirals require the integration of nine coupled ordinary differential equations (ODEs) for the components of $\mathbf{L}$, $\mathbf{S_1}$ and $\mathbf{S_2}$. Although the time $t$  at a given separation $r$ is crucial to calculate the emitted GW signal, it is not relevant for most astrophysical purposes, where only the evolution of the spin orientations is needed. For this reason, \precession\ performs PN integrations using the separation $r$ as independent variable. In practice, we integrate $d\mathbf{\mathcal L}/dr=d\mathbf{\mathcal L}/dt \times  (dv/dt)^{-1} \times 1/2\sqrt{r M}$, where $\mathcal L$ is any of the components of $\mathbf{L}$, $\mathbf{S_1}$ and $\mathbf{S_2}$. Integrations are performed using the  \comp{lsoda} algorithm  \citep{hindmarsh1982odepack}  implemented in \comp{scipy.integrate.odeint}. \comp{lsoda} combines adaptive nonstiff and stiff  methods and  monitors the ODE integrations to switch between the two as needed. 

We provide three convenient wrappers of the orbit-averaged PN integrator, which differ in their input and output parameters:
\begin{enumerate}
\item \fun{orbit\_averaged} evolves the relative orientation of the three momenta given in terms of $(\xi,J,S)$. The initial configurations must be compatible with the constraints presented in Sec.~\ref{geocon}.
\item \fun{orbit\_angles} evolves BH binary configurations specified by the angles $(\theta_1,\theta_2,\Delta\Phi)$.
\item \fun{orbit\_vectors} tracks the evolution of the nine components of $\mathbf{L}$ of $\mathbf{S_1}$ and $\mathbf{S_2}$ in an inertial frame.
\end{enumerate}
In the first two cases, the integration is carried out  in a reference frame $(\mathbf{\hat x},\mathbf{\hat y},\mathbf{\hat z})$  defined by $\mathbf{J}\cdot \mathbf{\hat x}=\mathbf{J}\cdot \mathbf{\hat y}=\mathbf{L}\cdot \mathbf{\hat y}=0$ at the initial separation; generic configurations can be projected to this frame using \fun{Jframe\_projection}. In the third case, the integration frame is specified by the input parameters. Examples are shown in Sec.~\ref{PNwrappers}.
Parallelization is implemented in all wrappers to evolve distributions of BH binaries on multiple cores (cf. Sec.~\ref{timing}). If needed, the right-hand side of Eqs.~(\ref{precession1})-(\ref{radiationreaction}) can be accessed explicitly calling \fun{orbav\_eqs}.

\subsection{Precession-averaged evolutions}
\label{precav}

References \cite{2015PhRvL.114h1103K,2015PhRvD..92f4016G} introduced an alternative way to evolve spinning BH binaries, which explicitly exploits the timescale hierarchy $t_{\rm pre}\ll t_{\rm RR}$. The three parameters $(\xi,J,S)$ describing the relative orientations of the BH spins naturally accommodate the timescales of the problem: 
\begin{itemize}
\item $\xi$ is conserved on both $t_{\rm pre}$ and $t_{\rm RR}$; 
\item $J$ is conserved on $t_{\rm pre}$ but varies on $t_{\rm RR}$; 
\item $S$ varies on both $t_{\rm pre}$ and $t_{\rm RR}$. 
\end{itemize}
The oscillations of $S$ on $t_{\rm pre}$  can be averaged over to study the binary evolution on times $t \sim t_{\rm RR}$. The secular  variation of $J$ on $t_{\rm RR}$ is given at 1PN by 
\begin{align}\label{dJdr}
\frac{dJ}{dr} 
= \frac{1}{4 r J} \left(J^2 + L^2- \frac{
\int_{S_-}^{S_+}   S^2 |dS/dt| ^{-1} dS}{\int_{S_-}^{S_+}  |dS/dt| ^{-1} dS} \right)~.
\end{align}
This approach reduces the PN evolution of a BH binary to a single ODE. The price paid to achieve this  simplification is the loss of information on the evolution of $S$ (cf. Sec.~\ref{phaseresamp} below). 

The integration domain of Eq.~(\ref{dJdr}) can be extended to arbitrarily large separations using auxiliary variables 
\begin{align}\label{kappadef}
\kappa= \frac{J^2 - L^2}{2 L}\,,\qquad  u=\frac{1}{2L}\,,
\end{align}
such that Eq.~(\ref{dJdr}) reduces to
\begin{align}\label{dkappadu}
\frac{d\kappa}{du} 
= \frac{
\int_{S_-}^{S_+}   S^2 |dS/dt| ^{-1} dS}{\int_{S_-}^{S_+}  |dS/dt| ^{-1} dS} ~,\end{align}
which can be integrated from/to $u=0$ ($r/M= \infty$). While $J\sim L \propto \sqrt{r}$ diverges in the large separation limit,  the asymptotic value  of $\kappa$,
\begin{align}
\kappa_\infty=\lim_{r/M\to\infty} \kappa = \lim_{r/M\to\infty} (\mathbf{S_1}+\mathbf{S_2})\cdot \hat{\mathbf{L}}
\end{align}
converges and becomes equivalent to the projection of the total spin along the orbital angular momentum.  $\kappa_\infty$ is, therefore, bounded by
\begin{align}
-(S_1+S_2) \,\leq \,&\kappa_{\infty}\, \leq S_1+S_2, \label{kappainflim}
\end{align}
as given by \fun{kappainf\_lim}.  BH binary configurations at infinitely large separation are  specified by pairs $(\xi,\kappa_\infty)$ satisfying Eqs.~(\ref{xilim}) and (\ref{kappainflim}); see Sec.~\ref{parselsec}. The allowed range of these two parameters can be computed using \fun{kappainf\_allowed} and \fun{xiinf\_allowed}. $\theta_1$ and $\theta_2$ asymptote to finite values at large separation, and can be expressed in terms of $\xi$ and $\kappa_\infty$:
\begin{align} 
\cos\theta_{1\infty}&\equiv\lim_{r/M \to \infty}  \cos\theta_1= \frac{ \kappa_\infty(1+q^{-1}) - M^2\xi }{S_1(q^{-1}-q)}\,,
\\
\cos\theta_{2\infty}&\equiv\lim_{r/M \to \infty}  \cos\theta_2 = \frac{M^2\xi - \kappa_\infty(1+q)}{S_2(q^{-1}-q)}\,.
\end{align}
Transformations between $(\xi,\kappa_\infty)$ and  $(\theta_{1\infty},\theta_{2\infty})$ are implemented in \fun{thetas\_inf} and \fun{from\_the\_angles\_inf}.

\precession\ provides three different wrappers to  integrate Eqs.~(\ref{dJdr}) and (\ref{dkappadu}):
\begin{enumerate}
\item \fun{evolve\_J} evolves the binary between two finite separations $r_i$ and $r_f$. The initial condition $J(r_i)$ must satisfy the geometrical constraints of Sec.~\ref{geocon}.
\item \fun{evolve\_J\_infinity} integrates Eq.~(\ref{dkappadu}) from $r/M = \infty$ ($u=0$) down to some final separation $r_f$. The initial configuration has to be specified in terms of  $\kappa_{\infty}$.
\item \fun{evolve\_J\_backwards} evolves a binary specified at some separation $r_i$ back to past infinity and returns its asymptotic condition $\kappa_\infty$.
\end{enumerate}
Practical examples are provided in Sec.~\ref{PNwrappers}. Integrations are performed using the  \comp{lsoda} algorithm \citep{hindmarsh1982odepack} wrapped in \comp{scipy.integrate.odeint}. Parallelization is implemented to run arrays of binaries simultaneously (cf. Sec.~\ref{timing}). The right-hand side of Eqs.~(\ref{dJdr}) and (\ref{dkappadu}) can be evaluated directly using \fun{dJdr} and \fun{dkappadu}.

When performing precession-averaged evolutions, we recommend avoiding binary configurations very close to the limits reported in Eqs.~(\ref{xilim})-(\ref{xiallowed}). Numerical errors arising from the integration of Eq.~(\ref{dJdr}) may push some of the parameters out of their  range of validity, which prevents any further evolution. \precession\ is rather solid with respect to such errors: tolerances as small as $\Delta J/M^2\sim\Delta \xi \sim 10^{-6}$ from the limits reported in Eqs.~(\ref{xilim})-(\ref{xiallowed}) are typically sufficient to ensure smooth integrations.

\subsection{Phase resampling and binary transfer}
\label{phaseresamp}

Precession-averaged integrations do not track the evolution of the precessional phase. This is a well justified approach for most astrophysical applications. Interactions with the astrophysical environment determine the spin orientation at  large separation where GW emission is inefficient to drive the dynamics \citep{2000ApJ...541..319K,2007ApJ...661L.147B,2008ApJ...682..474B,2010ApJ...719L..79F,2010MNRAS.402..682D,2013PhRvD..87j4028G,2015MNRAS.451.3941G,2013MNRAS.429L..30L,2013ApJ...774...43M,2014ApJ...794..104S}. The inequality $t_{\rm pre}\ll t_{\rm RR}$ implies that BH binaries undergo a very large number of precession cycles before entering the GW-driven regime, such that the information of the initial phase is lost in practice. 

An estimate of the magnitude of the total spin $S$ is nonetheless available at a statistical level from the  dynamics on the shorter times $t\sim t_{\rm pre}$.
The probability of finding a binary with some total spin magnitude $S$ is proportional to $dt/dS$  of Eq.~(\ref{dSdt}).
We sample the probability distribution $P(S)= 2 |dS/dt|^{-1} /\tau$ (with $S\in[S_-,S_+]$) using the cumulative distribution method (e.g. \cite{1997sda..book.....C}), which is suitable to handle  integrable singularities (cf. Eq.~\ref{integrablesings}). We first select a random number $\epsilon\in[0,1]$ and then solve the integral equation
\begin{align}
\frac{2}{\tau} \int_{S_-}^{S}   \frac{dS'}{|dS'/dt|}  = \epsilon
\end{align}
for $S\in[S_-,S_+]$. The algorithm is implemented in \fun{samplingS} and tested in Sec.~\ref{phaseres} below.

Phase resampling is essential to transfer the spin orientations of BH binaries from large separation where they form down to the regime close to merger. The complete procedure is implemented in \fun{evolve\_angles}, and can be summarized as follows.
\begin{enumerate}
\item We specify a binary with mass ratio $q$, spin magnitudes $S_1$, $S_2$ and spin orientations $(\theta_1,\theta_2,\Delta\Phi)$ at some initial separation $r_i$.
\item We convert the initial configuration to $(\xi,J,S)$ but only consider $(\xi,J)$, thus explicitly losing memory of $S$.
\item The configuration $(\xi,J)$ is evolved down to some final separation $r_f$ integrating Eq.~(\ref{dJdr}) for $J$ ($\xi$ stays constant).
\item Given the final configuration $(\xi,J)$ at $r_f$, we randomly extract a value $S$ from a distribution weighted by $|dS/dt|^{-1}$. 
\item The final set of parameters $(\xi,J,S)$  is converted back to $(\theta_1,\theta_2,\Delta\Phi)$. The sign of $\Delta\Phi$ is randomly chosen.
\end{enumerate}

This procedure allows for direct comparison between orbit-averaged and precession-averaged evolutions. Such a comparison  is carried out in Sec.~\ref{compevol} as a test of the code.
Tests performed on distributions of binaries have been reported by  \cite{2015PhRvD..92f4016G}, where precession-averaged binary transfers have been found to be in excellent statistical agreement with orbit-averaged evolutions. Discrepancies between the two approaches become relevant only at $r\sim 10 M$, where $t_{\rm pre}$ becomes comparable to $t_{\rm RR}$. However, the entire PN approach loses accuracy at such small separations \citep{2006PhRvD..74j4005B,2009PhRvD..80h4043B,2009PhRvD..79h4010C} and the binary evolution can be followed faithfully only using numerical-relativity simulations.

Neglecting and resampling the precessional phase lead to a substantial computational speed-up. A concrete example is provided in Sec.~\ref{timing}: even starting at  moderate separation $\sim 10^4M$, precession-averaged integrations are faster by about a factor $\sim 70$ when compared to orbit-averaged evolutions\footnote{Precession-averaged evolutions may occasionally stall and take longer to run. This is due to a wrong initial guess of the integration step
attempted by \comp{lsoda} and can be cured increasing the \comp{h0} optional parameter of \comp{scipy.integrate.odeint}. With the current default option, stalling happens roughly once every million inspirals.}. Orbit-averaged integrations become impractical at separations significantly larger than $\sim 10^4 M$, while precession-averaged evolutions can be carried out to/from infinitely large separation. 

\subsection{Hybrid evolutions}\label{hybrid}

Although optimal for statistical studies, phase resampling may be inaccurate in situations where individual precession cycles need to be resolved.  \precession\  can perform hybrid PN integrations combining the two approaches in \fun{hybrid}:
\begin{enumerate}
\item A precession-averaged integration is used at large separations, down to a certain separation threshold $r_t$. 
\item The precessional phase is extracted at $r_t$ by resampling the total spin magnitude $S$.
\item This binary configuration at $r_t$ is used to initialize an orbit-averaged integration to resolve individual precession cycles at separations $r<r_t$.
\end{enumerate}
The transition radius $r_t$ may correspond, for instance, to a typical separation  where the emitted GW frequency \mbox{$f_t=\sqrt{M/\pi^2 r_t^3}$} enters the lower end of the sensitivity window of a specific detector. For convenience, we provide utilities to convert binary separation and emitted GW frequency in \fun{rtof} and \fun{ftor}.

\section{Black-hole remnants}
\label{BHremnant}

\precession\ implements  numerical-relativity fitting formulas to estimate final mass (Sec.~\ref{secfinalmass}), spin (Sec.~\ref{secfinalspin}) and recoil (Sec.~\ref{secfinalkick}) of BHs following binary mergers. The importance of spin precession in estimating these properties is stressed in Sec.~\ref{impspin}.

The fitting formulas are typically written down using the following weighted combinations of the BH spin\footnote{Note the sign of $\mathbf{\Delta}$. We found inconsistencies in a few other publications, which we believe originate from converting expressions between notations where $m_2 \gtrless m_1$. We note it is a negligible effect to all practical purposes.}  \begin{align}
\mathbf{\Delta}&=\frac{\chi_1 \mathbf{\hat S_1} - q \chi_2 \mathbf{\hat S_2}}{1+q}\,,
\qquad
\mathbf{\tilde{\boldsymbol{\chi}}}=\frac{q^2 \chi_2 \mathbf{\hat S_2} + \chi_1 \mathbf{\hat S_1}}{(1+q)^2},
\label{chitilde}
\end{align}
and their projections parallel or perpendicular  to the orbital angular momentum:  
\mbox{$\tilde\chi_\parallel = \mathbf{\tilde{\boldsymbol{\chi}}}\cdot\mathbf{\hat L} $},
\mbox{$\tilde\chi_\perp = | \mathbf{\tilde{\boldsymbol{\chi}}}\times \mathbf{\hat L} |$},
\mbox{$\Delta_\parallel = \mathbf{\Delta}\cdot\mathbf{\hat L} $}, 
\mbox{$\Delta_\perp = | \mathbf{\Delta}\times \mathbf{\hat L} |$}.

\subsection{Final mass}
\label{secfinalmass}

The energy radiated in GWs during the inspiral and merger of a BH binary decreases the mass of the BH remnant $M_f$ below the binary's total mass $M$. Estimates of $M_f$ can be computed analytically in the test-particle limit $q\rightarrow 0$ \citep{2008PhRvD..78h4030K} and numerically in the strong-field regime $q \simeq 1$ \citep{2007PhRvD..76f4034B,2008PhRvD..78h1501T,2014PhRvD..89j4052L}. An interpolation between these two regimes is given in \cite{2012ApJ...758...63B} and reads
\begin{align}
\frac{M_f}{M}&=1 - \eta(1-4\eta) \left( 1- E_{\rm ISCO}\right)
\notag \\ &-16 \eta^{2} \left[p_0 +4 p_1 \tilde \chi_\parallel \left(\tilde \chi_\parallel +1\right) \right]\label{erad}\,.
\end{align}
Here $E_{\textsc{isco}}$ is the energy per unit mass of an effective particle of spin $\mathbf{\tilde{\boldsymbol{\chi}}}$  at the innermost stable circular orbit
\citep{1973blho.conf..215B}:
\begin{align}
E_{\textsc{isco}} &= \sqrt{1-\frac{2}{3 r_{\textsc{isco}}}}\,, \\
r_{\textsc{isco}}&=3+ Z_2
- {\rm sign}(\tilde \chi_\parallel)\sqrt{(3-Z_1)(3+Z_1+2Z_2)}\,,
\\
Z_1&=1+\left(1-\tilde \chi_\parallel^2\right)^{1/3}\left[\left(1+\tilde \chi_\parallel\right)^{1/3} + \left(1-\tilde \chi_\parallel\right)^{1/3}\right]\,,
\\
Z_2&=\sqrt{3\tilde \chi_\parallel^2 + Z_1^2}.
\end{align}
The parameters $p_0=0.04827$ and $p_1=0.01707$ have been obtained by \cite{2012ApJ...758...63B} fitting 186 numerical-relativity simulations from various groups. $M_f$ can be computed calling \fun{finalmass}.

\subsection{Final spin}
\label{secfinalspin}

A convenient expression for the spin $S_f=M_f^2 \chi_f$ of the BH remnant is given in \cite{2009ApJ...704L..40B}, where test-particle results \citep{2008PhRvD..77b6004B,2008PhRvD..78h4030K} and numerical-relativity simulations \citep{2008PhRvD..78h1501T,2008PhRvD..78d4002R,2014PhRvD..89j4052L} are interpolated. Their expression for the dimensionless spin $\chi_f$ is implemented in  \fun{finalspin} and reads 
\begin{align}
\chi_f&=\min\left(1, \left|\mathbf{\tilde{\boldsymbol{\chi}}} + \frac{q}{(1+q)^2}\ell\, \mathbf{\hat L}\right|\right) \label{finalspin}\,,
\\
\ell &= 2\sqrt{3} + t_2 \eta + t_3 \eta^2 + s_4 \frac{(1+q)^4}{(1+q^2)^2} \tilde \chi^2
\notag \\ &
+ (s_5 \eta + t_0 +2)\frac{(1+q)^2}{1+q^2} \tilde \chi_\parallel\,,
\end{align}
with $t_0=-2.8904$, $t_2=-3.51712$, $t_3=2.5763$, $s_4=-0.1229$ and $s_5=0.4537$. 

Various alternative prescriptions for the final spin have been compared in \cite{2010PhRvD..81h4054K}, where  the critical importance of accounting for PN spin precession in estimating $\chi_f$ is demonstrated (see also the discussion by \cite{2009ApJ...704L..40B} on this point).

\subsection{Black-hole recoil}
\label{secfinalkick}

If GWs are emitted anisotropically during  inspiral and merger, linear momentum is dissipated in a preferential direction and the center of mass recoils in the opposite direction. BH recoil (or kick) velocities $v_k$ can be as large as $\sim 5000$ km/s, which exceeds the escape velocities of the most massive galaxies \citep{2004ApJ...607L...9M}.  Kicks are generated by asymmetries in either the masses or the spins of the two merging BHs. The mass asymmetry contribution to the kick velocity $v_m $ lies in the orbital plane, while the spin contribution has components $v_{s\parallel}$ and $v_{s\perp}$ directed parallel and perpendicular to the orbital angular momentum. The magnitude of the kick velocity $v_k$ can be modeled as \citep{2007ApJ...659L...5C}
\begin{align}
\label{vkickall}
v_k= \sqrt{v_m^2 +2 v_m v_{s\perp} \cos\zeta + v^2_{s\perp} + v^2_{s\parallel}}\,,
\end{align}
where $\zeta$ is the angle between the mass term and the orbital-plane spin term. We have implemented the following expressions for $v_m$, $v_{s\perp}$ and $v_{s\parallel}$,
\begin{align}
v_m&= A \eta^2\frac{1-q}{1+q}(1+ B\eta)\,,  \\
v_{s\perp}&= H \eta^2 \Delta_\parallel\,, \\
v_{s\parallel}&=16 \eta^2  [ 
\Delta_\perp (V_{11} + 2 V_{ A} {\tilde \chi_\parallel}+4 V_{ B} {\tilde \chi_\parallel^2}+8 V_{ C} {\tilde \chi_\parallel^3}) 
\notag \\ &
+2 {\tilde \chi_{\perp}}\Delta_\parallel  (C_2 + 2 C_3 {\tilde \chi_{\parallel}})
] \cos{\Theta} \,,
\label{vparallel}
\end{align}
where the  coefficients are extracted from numerical-relativity simulations: $A = 1.2 \times 10^4~ {\rm km/s}$, $B = -0.93$ \citep{2007PhRvL..98i1101G}, $H=6.9  \times 10^3 ~{\rm km/s}$ \citep{2008PhRvD..77d4028L},  $V_{11} = 3677.76 ~{\rm km/s}$, $V_A  = 2481.21~ {\rm km/s}$, $V_B  = 1792.45 ~{\rm km/s}$, $V_C  = 1506.52 ~{\rm km/s}$ \citep{2012PhRvD..85h4015L}, $C_2=1140 ~{\rm km/s}$, $C_3=2481~ {\rm km/s}$ \citep{2013PhRvD..87h4027L}, $\zeta=145^\circ$ \citep{2008PhRvD..77d4028L}.  The main contribution to $v_k$   comes from the term proportional to $V_{11}$ in Eq.~(\ref{vparallel}). This effect, known as ``{superkick},'' enters $v_k$ weighted by $\Delta_\perp$ and it is dominant if binaries merge with $\theta_i\sim \pi/2$ and $\Delta\Phi\sim \pi$ 
\citep{2007PhRvL..98w1101G,2007ApJ...659L...5C}. The additional corrections $V_{A,B,C}$ ($C_{2,3}$) are known as ``{hangup-kicks}'' (``cross-kicks'')  and increase $v_k$ for moderate misalignments $\theta_i\sim 50^\circ$  \citep{2011PhRvL.107w1102L,2013PhRvD..87h4027L}. The additional parameter $\Theta$  is the angle between the direction of $ \mathbf{\Delta}\times \mathbf{\hat L}$ and the infall direction of the two holes ``at merger," offset by $\sim 200^\circ$ \citep{2008PhRvD..77l4047B,2009PhRvD..79f4018L}. In practice, $\Theta$ depends on the initial separation of the BH binary in each  numerical-relativity simulation. Following  previous studies \citep{2012PhRvD..85h4015L,2012PhRvD..85l4049B,2016MNRAS.456..961B}, \precession\  deals with this dependency assuming $\Theta$  to be uniformly distributed in $[0,\pi]$. Possible PN effects on the probability distribution of $\Theta$ are not taken into account.

Equation~(\ref{vkickall}) can be evaluated using \fun{finalkick} and predicts a maximum kick velocity  $v_k\sim 0.017 c \sim 5000$ km/s.

\subsection{Importance of spin precession}
\label{impspin}

Spin precession plays a crucial role in determining the properties of the BH remnant.  The fitting formulas here presented should only be applied at separations $r\lesssim {10} M$ comparable to the initial conditions of the numerical-relativity simulations used in their calibration. The PN inspiral before merger profoundly modifies the spin orientations and therefore the estimated properties of the final BH. Reference \citep{2010PhRvD..81h4054K} showed that PN spin precession introduces a fundamental uncertainty in predicting the final spin because $\chi_f$ depends on the precessional phase at merger, which is only available at a statistical level. This point is even more crucial for kick predictions. Large kicks are expected to be less (more) likely if binaries merge with $\Delta\Phi\sim0$ ($\sim \pi$) \citep{2007PhRvL..98w1101G,2007ApJ...659L...5C} and, consequently, phase transitions towards the librating morphologies during the early inspiral  substantially  suppress (enhance) the recoil \citep{2010ApJ...715.1006K,2012PhRvD..85l4049B}. Reference~\cite{2015PhRvD..92f4016G} found that binary morphologies close to merger are closely related to the spin configurations at large separations, which opens up the possibility of exploiting future BH kick measurements to constrain the astrophysical  processes behind BH binary formation and evolution \citep{2008ApJ...684..822B,2010MNRAS.404.2143V,2012AdAst2012E..14K,2013MNRAS.429L..30L,2015MNRAS.446...38G,2016MNRAS.456..961B,2016ApJ...818L..22A}. 

The expressions for final mass, spin and recoil currently implemented in \precession\ are the same already presented in \cite{2015MNRAS.446...38G}; other recent findings (e.g. \cite{2014PhRvD..89j4052L,2014PhRvD..90j4004H,2015PhRvD..92b4022Z,2016PhRvD..93d4006H})  will be implemented in future versions of the code.

\section{Examples} 
\label{examples}

This section contains several practical examples  for using  \precession. All tests presented here are available in the \textsc{python} submodule \fun{precession.test} which  has to be loaded explicitly with the command:
\begin{verbatim}
    import precession.test
\end{verbatim}
The source code of the example routines are reported in Figs.~\ref{parselcode}-\ref{timingcode}. The outcome of their executions are presented as screen outputs or graphical plots in Figs.~\ref{parselout}-\ref{timingout}. Each example is described in a dedicated subsection:
Sec.~\ref{parselsec} shows how to select consistent BH binary configurations and study their dynamics on $t_{\rm pre}$; in Sec.~\ref{spansec}, we study the precessional cycles of three BH binaries and classify their spin morphologies; in Sec.~\ref{phaseres}, we test our algorithm to resample the precessional phase; Sec.~\ref{PNwrappers} shows how to compute PN inspirals and  evaluate numerical-relativity fitting formulas to estimate the properties of the postmerger BH; finally, in Sec.~\ref{compevol} and \ref{timing} we compare  binary dynamics and computational speed of orbit-averaged and precession-averaged integrations. 

\subsection{Selection of consistent parameters}
\label{parselsec}

\begin{figure*}[p]
\includegraphics[width=\textwidth]{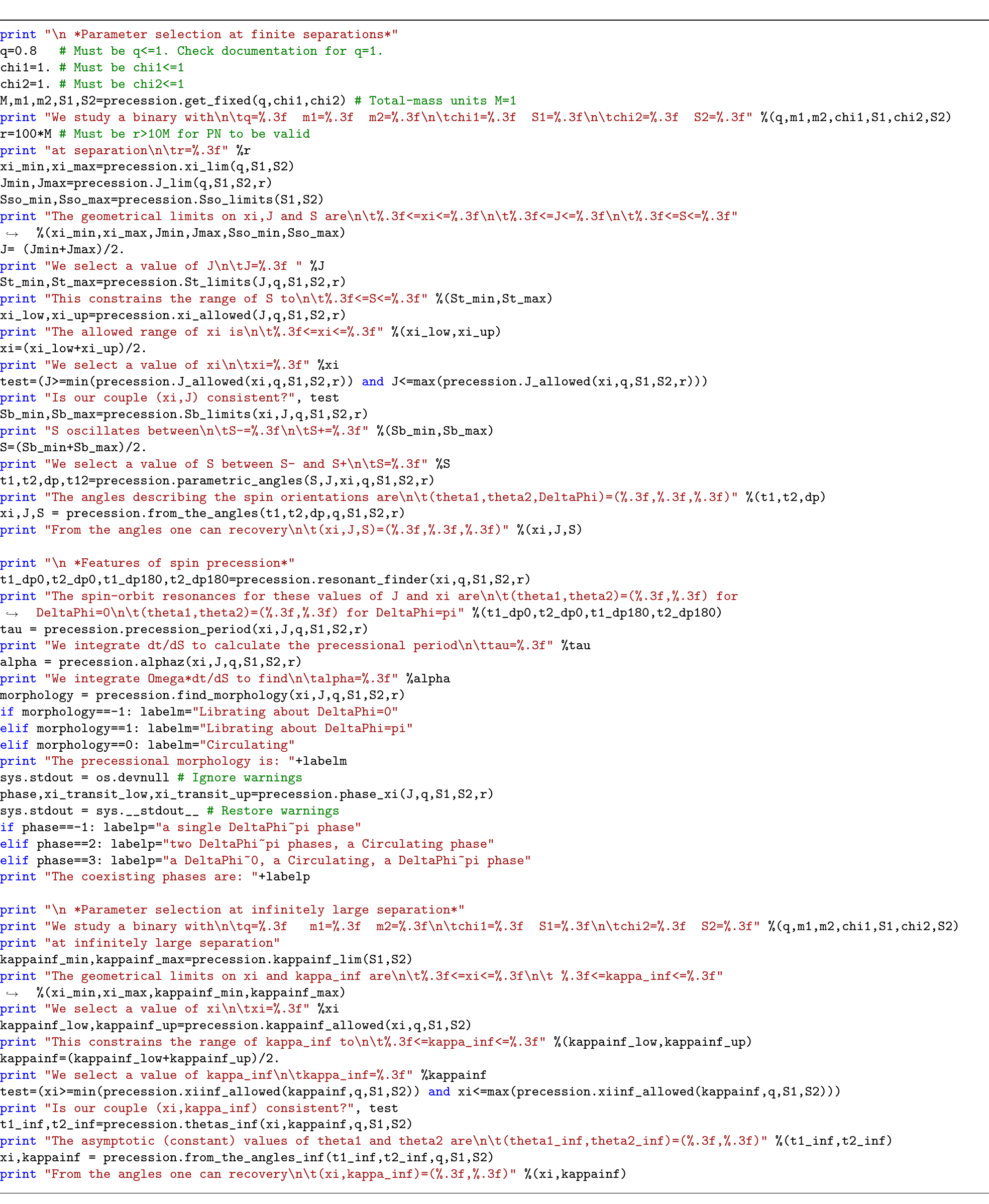}

\caption{Source code of \fun{test.parameter\_selection}, described in Sec.~\ref{parselsec}. The screen output is reported in Fig.~\ref{parselout}. In this example we (i) select consistent parameters at finite separation, (ii) compute several quantities to characterize the precessional dynamics and (iii) select consistent parameters at infinitely large separation. This test is run typing \comp{precession.test.parameter\_selection()}. \label{parselcode}}
\end{figure*}

The  function \fun{test.parameter\_selection} illustrates how to select consistent parameters and characterize the binary dynamics on the precession timescale. The test is executed with
\begin{verbatim}
    precession.test.parameter_selection()
\end{verbatim}
The source code and the screen output are reported in Figs.~\ref{parselcode} and \ref{parselout} respectively. 

We first show how to select values of $(\xi, J, S)$ that satisfy the geometrical constraints described in Sec.~\ref{geocon}, and how to convert these values to $(\theta_1,\theta_2,\Delta\Phi)$. Secondly, we compute several quantities that characterize BH spin precession: the angles $\theta_i$ corresponding to the spin-orbit resonances, the precessional period $\tau$, the total precession rate $\alpha$ and the spin morphology. Finally, we illustrate how to select consistent parameters at infinitely large separation $r/M\to \infty$ (cf. Sec.~\ref{precav}).

\subsection{Evolutions of the spin angles on a precession cycle}
\label{spansec}

\begin{figure*}[p]
\includegraphics[width=\textwidth]{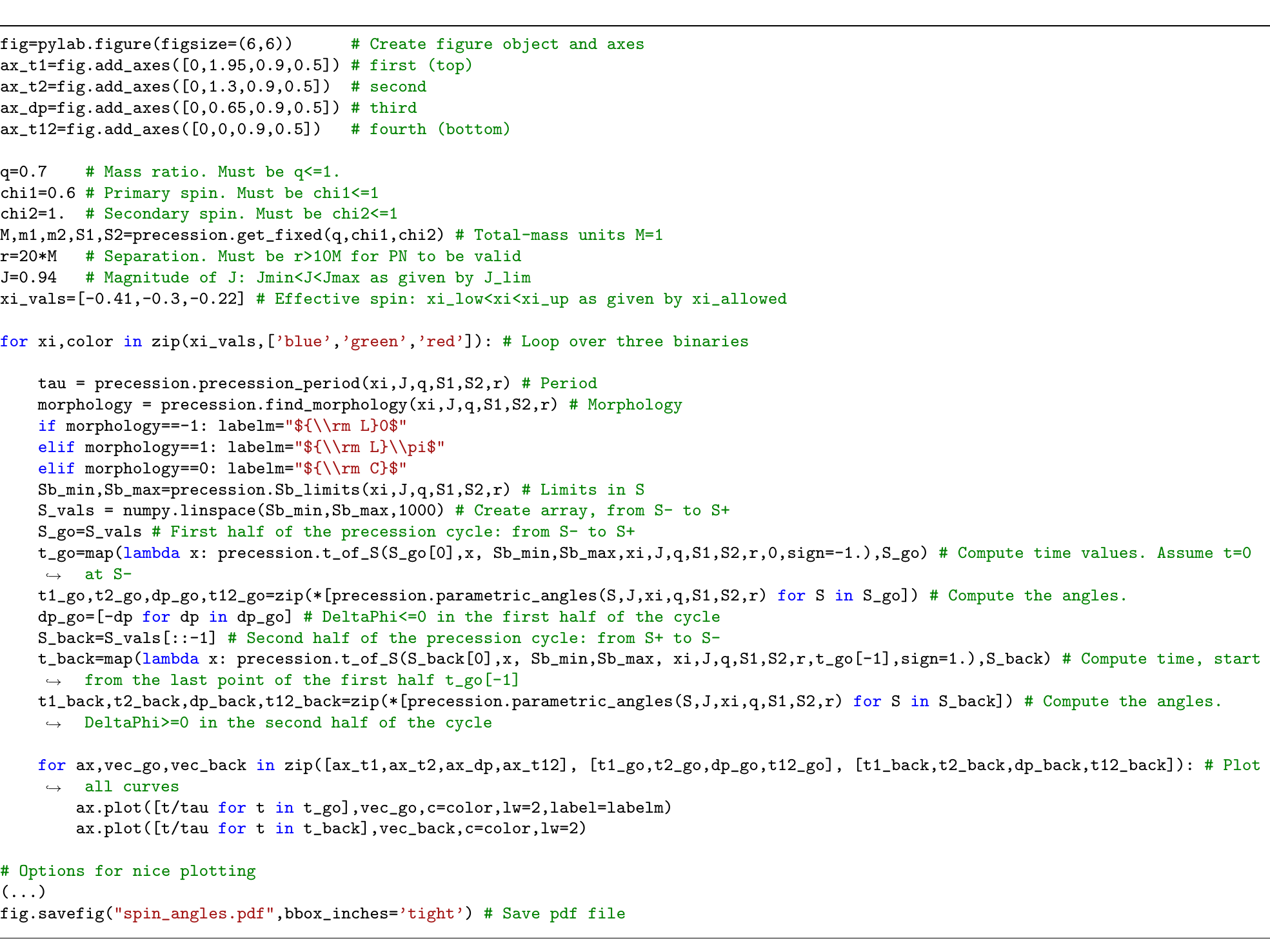}
\caption{Source code of \fun{test.spin\_angles}, described in Sec.~\ref{spansec}. The resulting plot is shown in  Fig.~\ref{spanout}. This example illustrates how to study the evolution of the angles $\theta_1$, $\theta_2$, $\Delta\Phi$ and $\theta_{12}$ over a single precession cycle $S_- \to S_+ \to S_-$. The precessional dynamics is first parametrized using $S$, and then plotted in terms of the time $t$ integrating $dS/dt$ from Eq.~(\ref{dSdt}). We assume $S=S-$ at $t=0$ and match the two halves of the precession cycle at $S=S+$. Note that the sign of $\Delta\Phi$ has to be specified by the user. Three binaries are considered here; their precessional morphology is evaluated and used to fill the plot legend.  This test is run typing \comp{precession.test.spin\_angles()}. Additional plotting options present in the source code have been omitted. }
\label{spancode}
\end{figure*}

The function \fun{test.spin\_angles} provides an example to study the binary evolution over one  single precession cycle. The test is executed with
\begin{verbatim}
    precession.test.spin_angles()
\end{verbatim}
The source code is reported in Fig.~\ref{spancode}; the resulting plot is shown in Fig.~\ref{spanout}.

The  separation $r$ and the magnitude of the total angular momentum $J$ are approximately constant on times $t\sim t_{\rm pre}$. Combined with the conservation of $\xi$, this implies that the precessional dynamics can be parametrized using a single parameter. We first parametrize the precession cycles using the magnitude of the total spin $S$. Time evolutions are then obtained by integrating $dS/dt$ according to Eq.~(\ref{tofS}). The magnitude $S$ undergoes a full oscillation between two values $S_-$ and $S_+$ in a time $\tau$ [cf. Eq.~(\ref{tau})], which defines the precession period.
As shown in Fig.~\ref{spanout},
the evolution of the tilt angles $\theta_i$ is qualitatively similar for all binaries. On the other hand, three different situations are possible for $\Delta\Phi$ and exemplify the notion of precessional morphology. As already pointed out in Sec.~\ref{params}, the sign of $\Delta\Phi$ must be specified by the user: one has $\Delta\Phi\leq 0$ ($\Delta\Phi\geq 0$) in the first (second) half of the precession cycle where $S$ increases (decreases).

\subsection{Sampling of the precessional phase}
\label{phaseres}

\begin{figure*}[p]
\includegraphics[width=\textwidth]{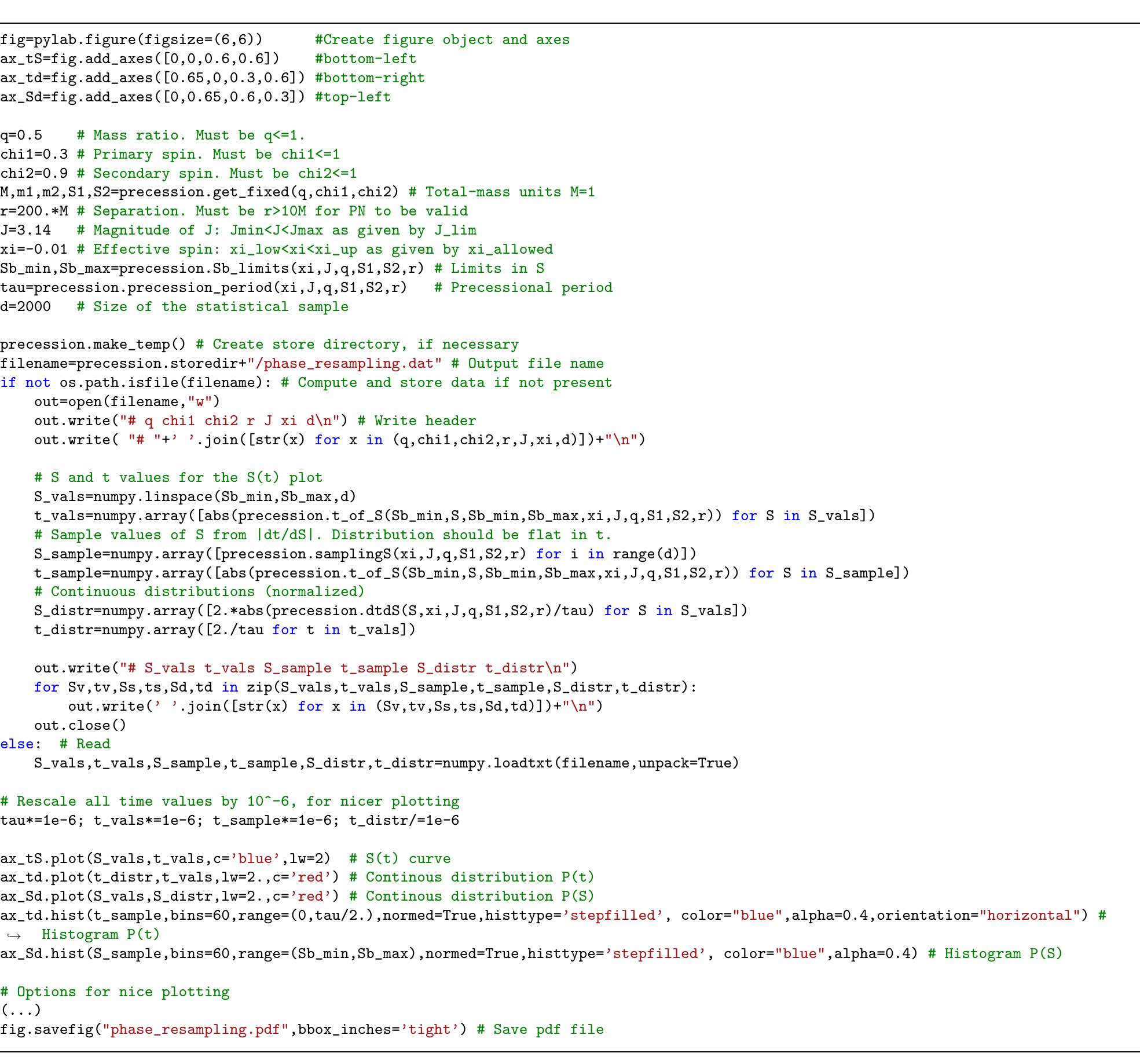}
\caption{Source code of \fun{test.phase\_resampling}, described in Sec.~\ref{phaseres}. The resulting plot is shown in Fig.~\ref{phaseresout}. We extract N=2000 values of the precessional phase $S$ from the probability distribution $P(S)= 2 |dS/dt|^{-1} /\tau$ in $[S-,S+]$. The procedure is illustrated in Sec.~\ref{phaseresamp} and is a key step to perform precession-averaged inspirals. We verify that the distribution $t(S)$ constructed  from the sampled values of $S$ is uniform in $[0,\tau/2]$. This test is run typing {\tt precession.test.phase\_resampling()}. Data are stored in \fun{precession.storedir}.  Additional plotting options present in the source code have been omitted. }
\label{phaserescode}
\end{figure*}

The  routine \fun{test.phase\_sampling} tests our procedure to statistically sample values of $S$ weighted by $|dS/dt|^{-1}$ (cf. Sec.~\ref{phaseresamp}). The test is executed with
\begin{verbatim}
    precession.test.phase_sampling()
\end{verbatim}
The source code is reported in Fig.~\ref{phaserescode}; the resulting plot is shown in Fig.~\ref{phaseresout}.

 After selecting a BH binary configuration $(q,\chi_1,\chi_2,r,J,\xi)$, we extract multiple values of $S\in[S_-,S_+]$ using \fun{samplingS}. The obtained distribution is normalized, binned, and compared with the continuum limit $P(S)= 2 |dS/dt|^{-1} /\tau$. As a consistency check, we also convert our sample to $t(S)$ using Eq.~(\ref{tofS}) and verify that these values are distributed uniformly.  This example also demonstrates that the singularities of $P(S)$  at $S_{\pm}$ [cf. Eq.~(\ref{integrablesings})] are integrable and result in a smooth probability distribution of $t$. 

\subsection{Wrappers of the PN integrators}
\label{PNwrappers}

\begin{figure*}[p]
\includegraphics[width=\textwidth]{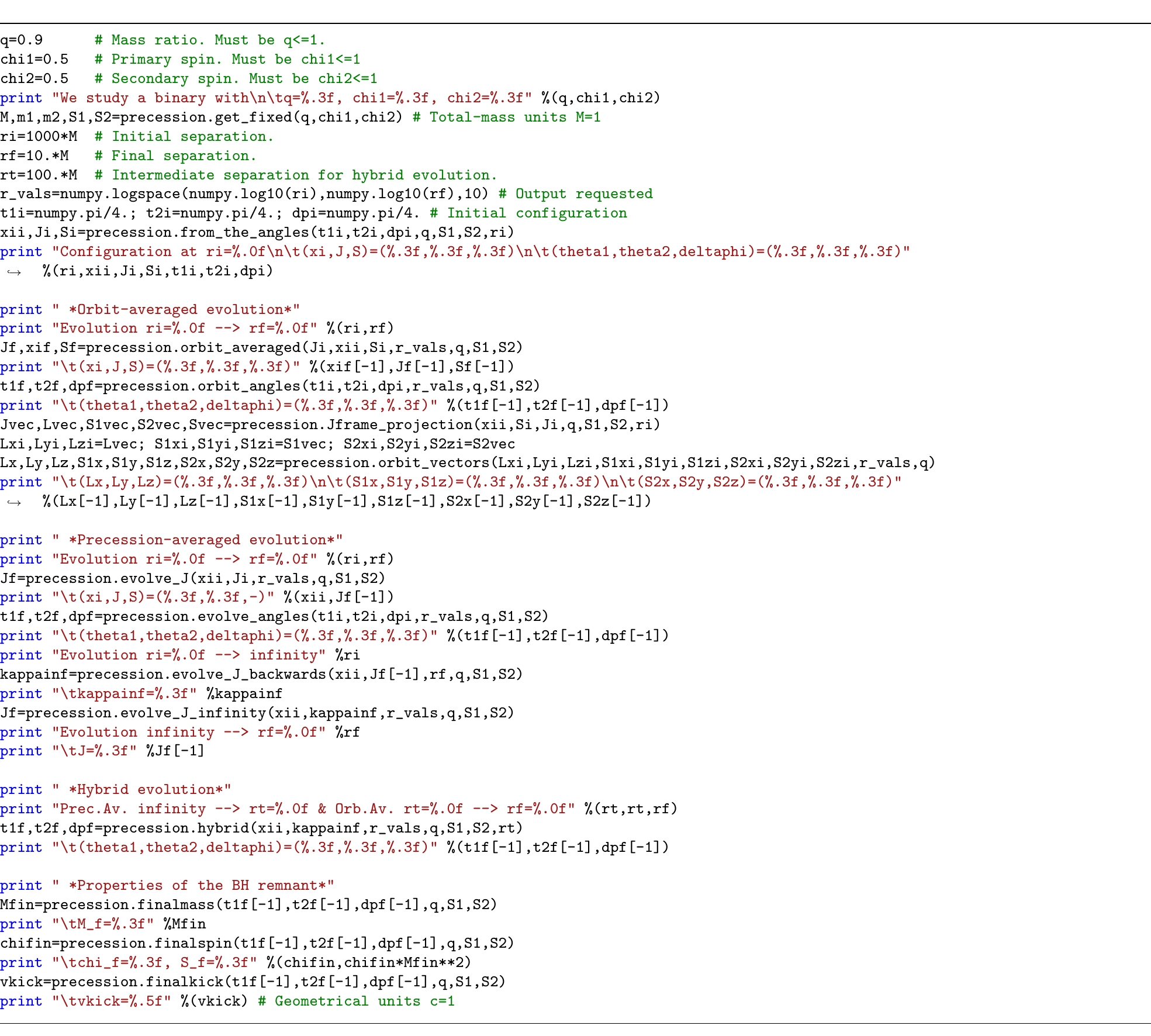}
\caption{Source code of \fun{test.PNwrappers}, described in Sec.~\ref{PNwrappers}. The screen output is reported in Fig.~\ref{pnwrapout}. This example shows how to use the various routines to perform PN inspiral. After specifying a BH binary at $r_i$, we evolve it down to $r_f$ using both orbit-averaged and precession-averaged integrations. We then  extract the asymptotic configuration $\kappa_\infty$ and  show how to match precession-averaged and orbit-averaged evolutions to construct hybrid inspirals. We also estimate mass, spin and recoil of the postmerger BH. This test is run typing \comp{precession.test.PNwrappers()}. Data are stored in the directory specified through \fun{precession.storedir}.}
\label{pnwrapcode}
\end{figure*}

The example \fun{test.PNwrappers} shows how to perform PN inspirals using the ODE integrators implemented in \precession. 
 The test is executed with
\begin{verbatim}
    precession.test.PNwrappers()
\end{verbatim}
The source code and the screen output are reported in Figs.~\ref{pnwrapcode} and \ref{pnwrapout}, respectively.

 We first specify a BH binary at some initial separation $r_i$ by providing values of the angles $(\theta_1,\theta_2,\Delta\Phi)$, which are then converted to $(\xi,J,S)$, cf. Sec.~\ref{params}. This system is first evolved down to a final separation $r_f<r_i$ integrating the orbit-averaged PN equations of motion (\ref{precession1})-(\ref{radiationreaction}). The integration is performed using the three wrappers presented in Sec.~\ref{orbav} to extract the final configuration in terms of $(\theta_1,\theta_2,\Delta\Phi)$, $(\xi,J,S)$, and the nine components of $\mathbf{L}$, $\mathbf{S_1}$, $\mathbf{S_2}$.
The same evolution is then performed using the precession-averaged approach outlined in  Sec.~\ref{precav}. In contrast to orbit-averaged integrations, $\xi$ is not evolved explicitly and it is assumed to be constant. The final value of $J$ is obtained by integrating  Eq.~(\ref{dJdr}). The evolution of $S$ is not tracked explicitly, but can be resampled (cf. Sec.~\ref{phaseresamp}) to obtain a statistical estimate of the angles $(\theta_1,\theta_2,\Delta\Phi)$ at $r_f$.
We then show how to perform integrations to/from $r/M\to\infty$, where the  projection of the total spin $\kappa_\infty$ is asymptotically constant. Finally, we evolve the same BH binary using a hybrid approach, stitching together precession-averaged and orbit-averaged integrations at some separation $r_t$. We complete this exercise with the evaluation of the numerical-relativity fitting formulas to estimate the properties of the postmerger BH remnant (Sec.~\ref{BHremnant}). Formulas are applied at $r_f$, \emph{after} the PN evolution.

\subsection{Comparison between orbit-averaged and precession-averaged integrations}
\label{compevol}

\begin{figure*}[p]
\includegraphics[width=\textwidth]{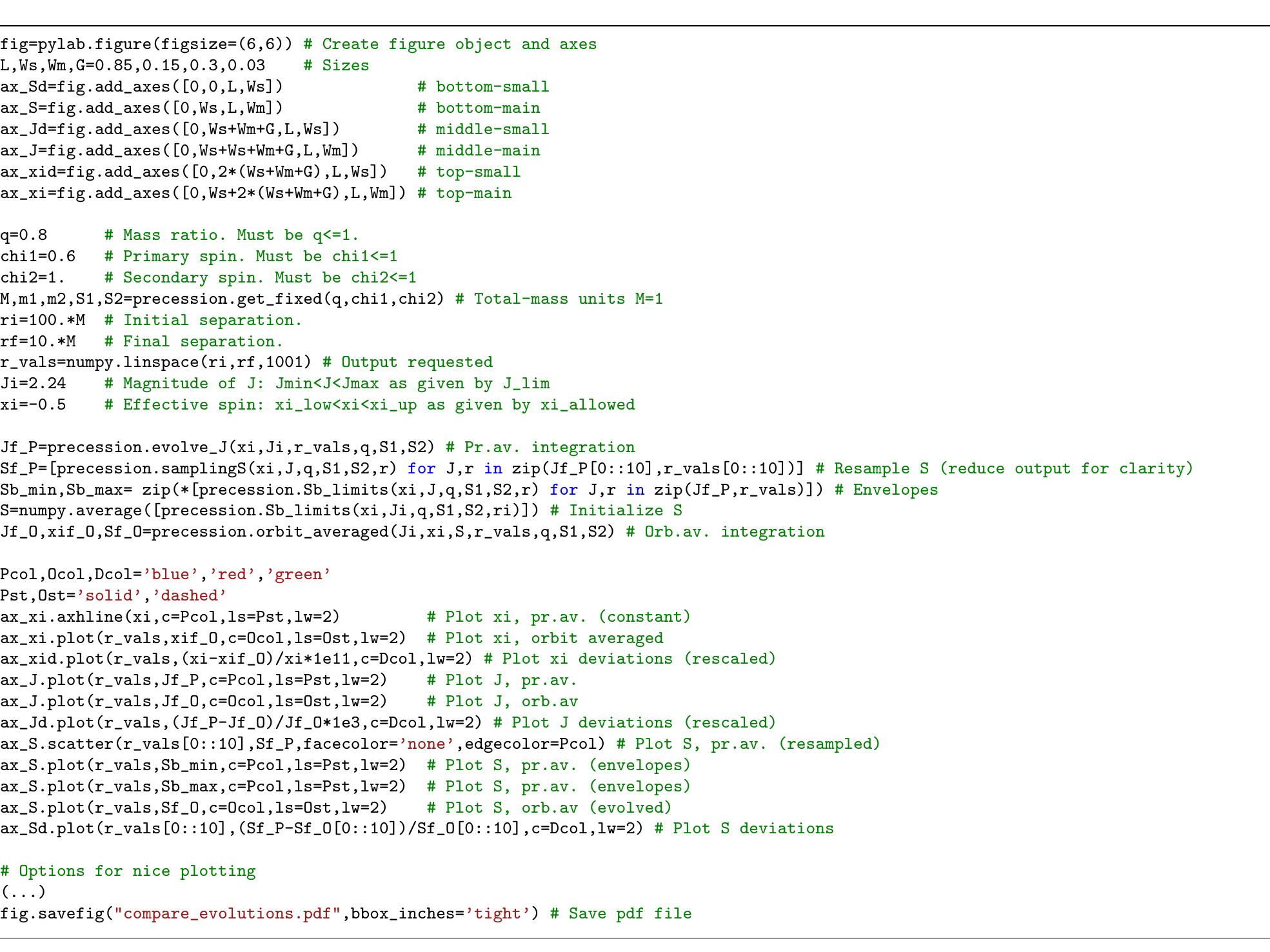}
\caption{
Source code of \fun{test.compare\_evolutions}, described in Sec.~\ref{compevol}. The resulting plot is shown in Fig.~\ref{compevolout}.  We compare precession-averaged and orbit-averaged integrations of a single BH binary. We  perform the two integrations from $r_i=100M$ to $r_f=10M$ and extract values of $\xi$, $J$ and $S$ along the inspiral. Relative differences between the two approaches are computed and plotted as a function of the binary separation. This test is run typing  \comp{precession.test.compare\_evolutions()}. Data are stored in the directory specified through \fun{precession.storedir}.  Additional plotting options present in the source code have been omitted. }
\label{compevolcode}
\end{figure*}

The example \fun{test.compare\_evolutions} compares a single PN evolution performed using orbit-averaged and precession-averaged integrations. The test is executed with
\begin{verbatim}
    precession.test.compare_evolutions()
\end{verbatim}
The source code is reported in Fig.~\ref{compevolcode}; the resulting plot is shown in Fig.~\ref{compevolout}.

Conservation of the effective spin $\xi$ on the precessional time  \citep{2001PhRvD..64l4013D,2008PhRvD..78d4021R} is a crucial assumption underlying our precession-averaged approach. On the other hand, orbit-averaged integrations confirm this feature as a byproduct. We detect extremely small  deviations $\Delta\xi/\xi\sim 10^{-11}$ between the two approaches (cf. top panel of Fig.~\ref{compevolout}), which fully corroborates our assumption, at least at the PN order  we implemented. Variations of $\xi$ due to additional PN corrections \citep{2013CQGra..30e5007M,2013CQGra..30g5017B,2015CQGra..32s5010B} still need to be explored. Note that $\xi$ is \emph{not} conserved on the orbital timescale (only on $t_{\rm pre}$ and $t_{\rm RR}$), but those variations are not captured by either of our methods.
 The evolution of $J$ is also very accurate, with deviations of the order of $\Delta J/ J\sim 10^{-3}$ during the entire integration (Fig.~\ref{compevolout}, middle panel).  The precession-averaged approach gradually loses accuracy at small separations, where the precession time $t_{\rm pre}$ becomes  comparable to the inspiral time $t_{\rm RR}$. Precession-averaged integrations require a resampling of the precessional phase $S$ at each output separation. Resampled values are in excellent \emph{statistical} agreement with the orbit-averaged result (lower panel of Fig.~\ref{compevolout}). The envelope of the orbit-averaged evolution of $S$ is  well described by the $S_\pm$ curves given by $\xi_{\pm}(S)=\xi$, cf. Eq.~(\ref{effpot}).

\subsection{Parallel computation and timing}
\label{timing}
\begin{figure*}[p]
\includegraphics[width=\textwidth]{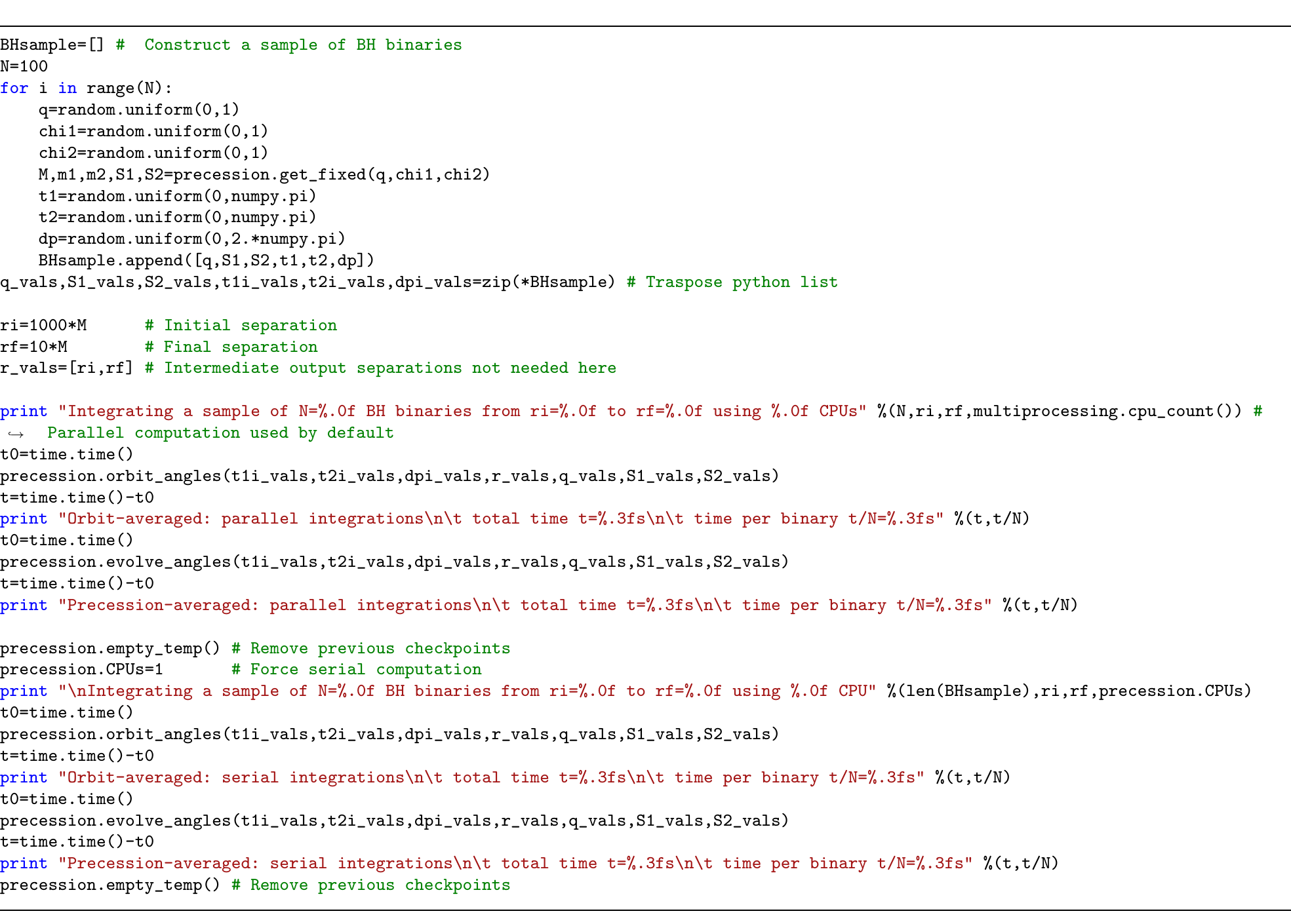}
\caption{Source code of \fun{test.timing} described in Sec.~\ref{timing}; the screen output is reported in Fig.~\ref{timingout}. We compute the CPU time needed to evolve a sample of $N=100$ binaries from $r_i=10^4M$ to $r_f=10M$ using both orbit-averaged and precession-averaged integrations. By default, \precession\  performs PN inspirals in parallel using all available cores. The two computations are repeated enforcing a strictly serial execution.  This test is run typing \comp{precession.test.timing()}. Data are stored in the directory specified through \fun{precession.storedir}.}
\label{timingcode}
\end{figure*}

\begin{figure*}[p]
\begin{minipage}[b]{0.99\columnwidth}
\includegraphics[width=0.99\columnwidth]{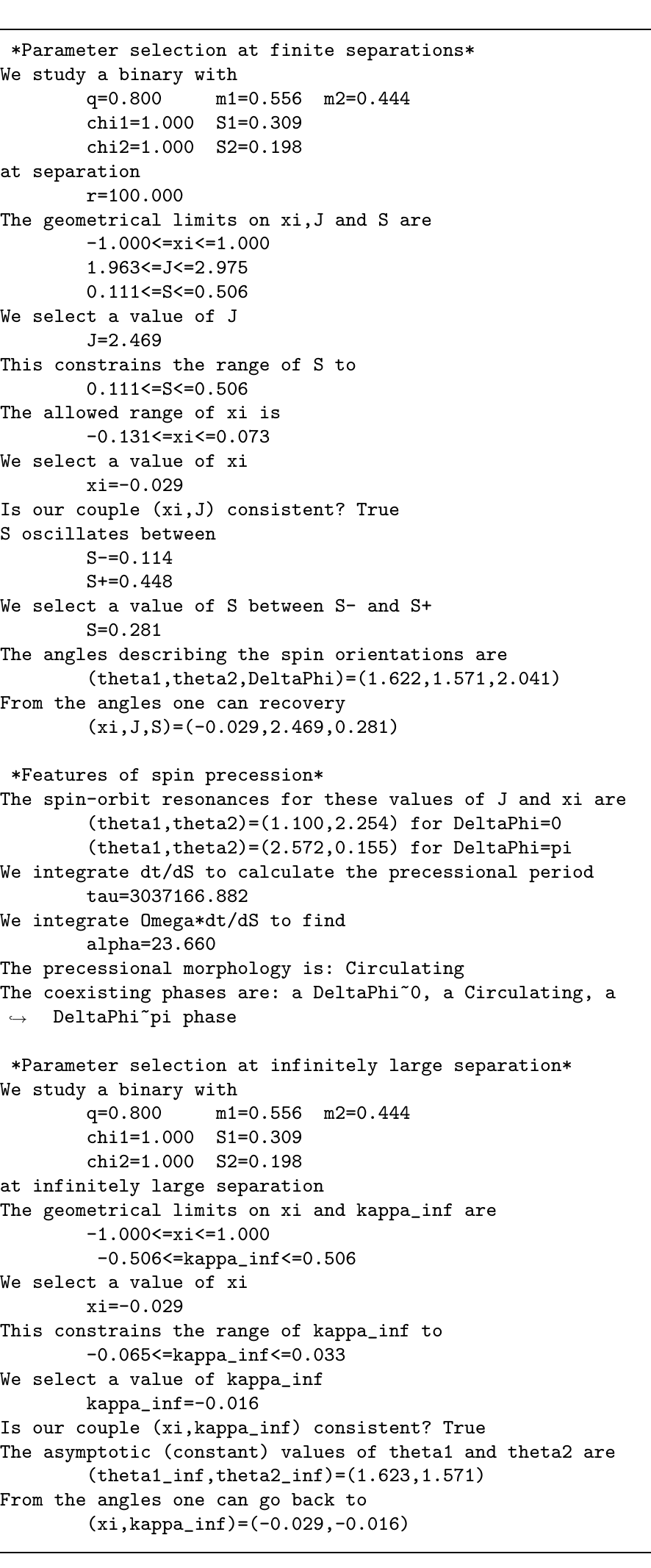}
\caption{Screen output of \fun{test.parameter\_selection}, described in Sec.~\ref{parselsec}. The source code is reported in Fig.~\ref{parselcode}. In this example we (i) select consistent parameters at finite separation, (ii) compute several quantities to characterize the precessional dynamics and (iii) select consistent parameters at infinitely large separation. Outputs have been rounded to three decimal digits for clarity.  This test is run typing \comp{precession.test.parameter\_selection()}.}
\vspace{0.9cm}
 \label{parselout}
\end{minipage}
\hfill
\begin{minipage}[b]{0.99\columnwidth}
\includegraphics[width=0.95\columnwidth]{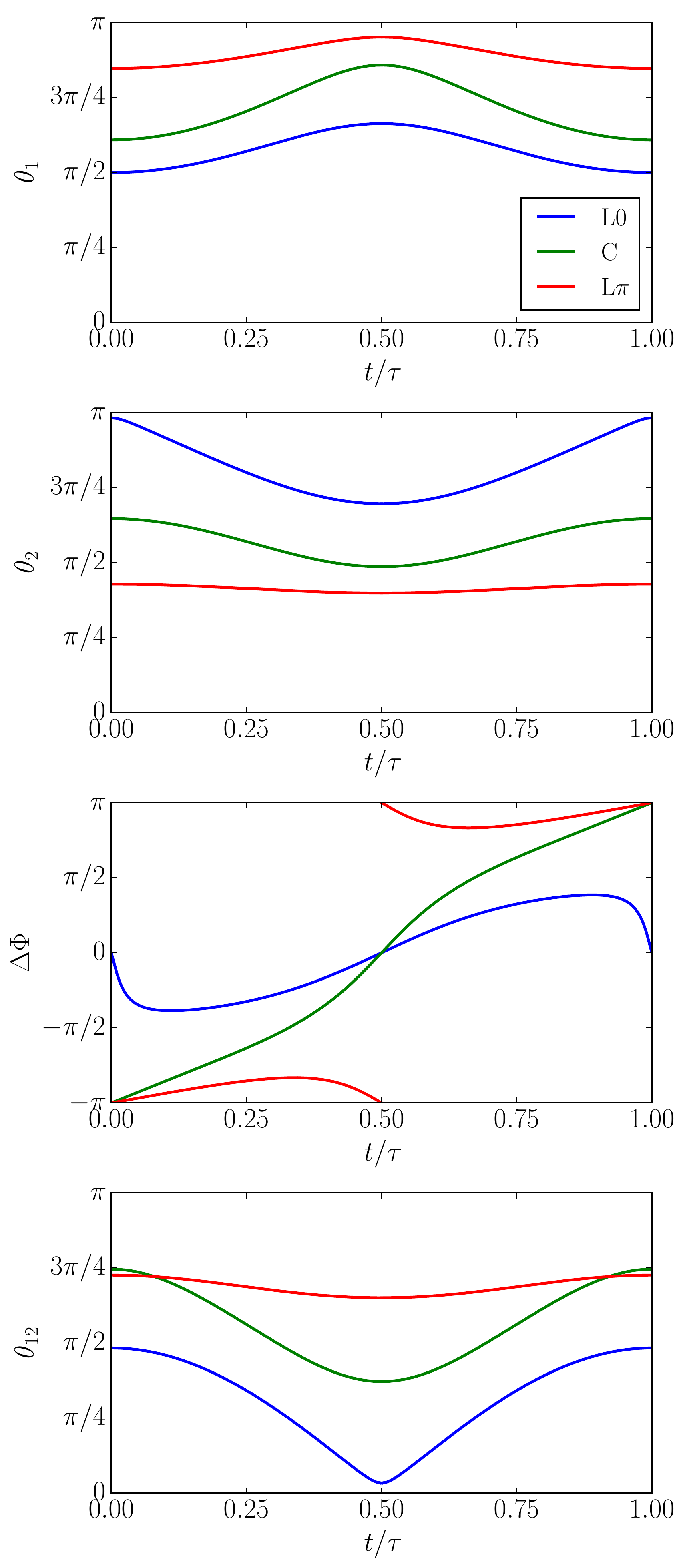}
\caption{Resulting plot obtained from \fun{test.spin\_angles}, described in Sec.~\ref{spansec}. The source code is reported in Fig.~\ref{spancode}. We study the precessional dynamics of  three binary BHs with mass ratio $q=0.7$, dimensionless spin $\chi_1=0.6$, $\chi_2=1$, total angular momentum $J=0.94 M^2$ at separation $r=20 M$. The evolution of the angles  $\theta_1$, $\theta_2$, $\Delta\Phi$ and $\theta_{12}$ (top to bottom) is plotted against  the time $t$ normalized to the precessional period $\tau$.  The configurations shown here are characterized by different values of the effective spin $\xi$ and belong to the three different morphologies: the binary with $\xi=-0.41$ (blue) is librating about $\Delta\Phi=0$ (L$0$); the  binary with  $\xi=-0.3$ (green) is circulating through the full range $\Delta\Phi\in[-\pi,\pi]$ (C), and  the binary with $\xi=-0.22$ (red) is librating about $\Delta\Phi=\pm\pi$ (L$\pi$). This test is run typing {\tt precession.test.spin\_angles()}.
}\label{spanout}
\end{minipage}
\end{figure*}

\begin{figure*}[p]
\begin{minipage}[b]{0.99\columnwidth}
\includegraphics[width=\columnwidth]{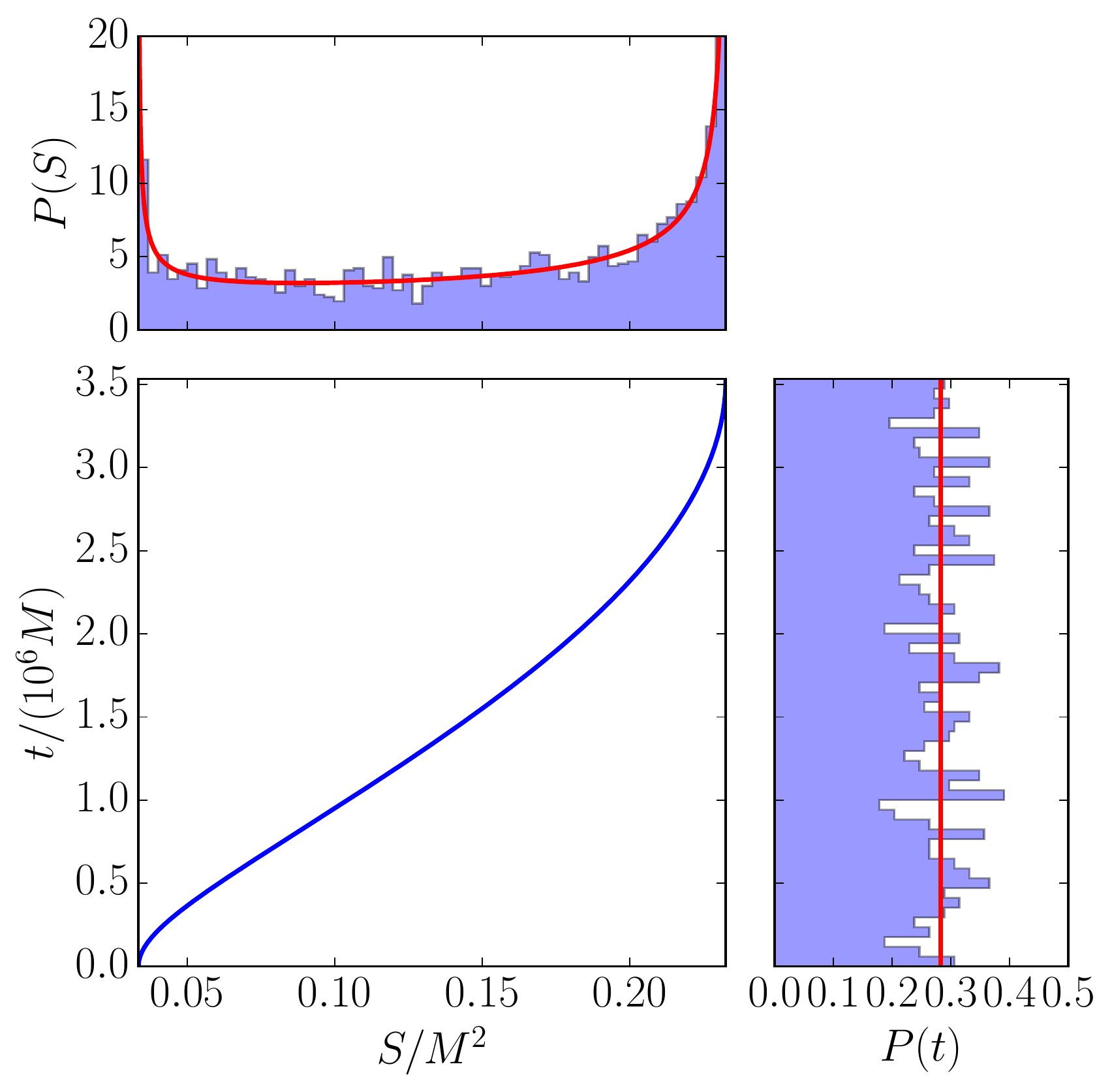}
\caption{Resulting plot obtained from \fun{test.phase\_resampling}, described in Sec.~\ref{phaseres}. The source code is reported in Fig.~\ref{phaserescode}. The bottom left panel shows the evolution of S on the precession time for a BH binary with $q=0.5$, $\chi_1=0.3$, $\chi_2=0.9$, $J=3.14 M^2$, $\xi=-0.01 $ and $r=200 M$. The binary evolves from $S_- \simeq 0.033$ ($t=0$) to $S_+ \simeq 0.232$ ($t=\tau/2\simeq 3.53\times 10^6 M$). We extract a sample of $N=2000$ values of $S$ from a probability distribution proportional to $|dS/dt|^{-1}$. Histograms of the extracted distribution of $S$  and $t$ are shown in the top and right panels, respectively, where red lines mark the  continuum  limit. This procedure efficiently extracts BH binaries according to their time spent at each  spin configuration and demonstrates the correct handling of the (integrable) singularities of $|dS/dt|^{-1}$ at $S_\pm$. This test is run typing \comp{precession.test.phase\_resampling()}.
}\label{phaseresout}
\end{minipage}
\hfill
\begin{minipage}[b]{0.99\columnwidth}
\includegraphics{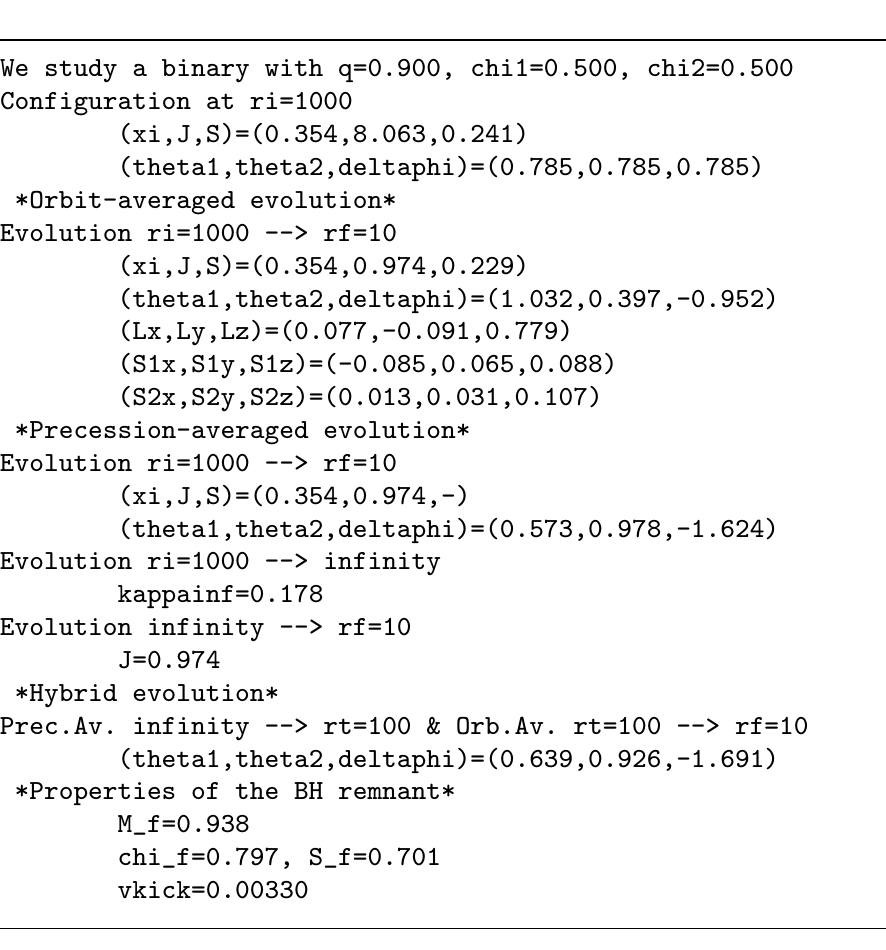}
\caption{Screen output of \fun{test.PNwrappers}, described in Sec.~\ref{PNwrappers}. The source code is reported in Fig.~\ref{pnwrapcode}.  After selecting a binary at the initial separation $r_i$, we (i) perform orbit-averaged integrations from $r_i$ to a final separation $r_f$; (ii) perform precession-averaged integrations from $r_i$ to $r_f$, from $r_i$ to $r/M=\infty$ and from $r/M=\infty$ to $r_f$, (iii) perform hybrid integrations from $r/M=\infty$ to $r_f$ matched at a separation threshold $r_t$ and (iv)  extract the properties of the BH remnant applying fitting formulas at $r_f$. Outputs have been rounded to three decimal digits for clarity;  output lines regarding the location of the stored data files have been omitted. This test is run typing \comp{precession.test.PNwrappers()}. \label{pnwrapout}}
\vspace{0.2cm}
\end{minipage}
\end{figure*}

\begin{figure*}[p]
\begin{minipage}[b]{0.99\columnwidth}
\includegraphics[width=\columnwidth]{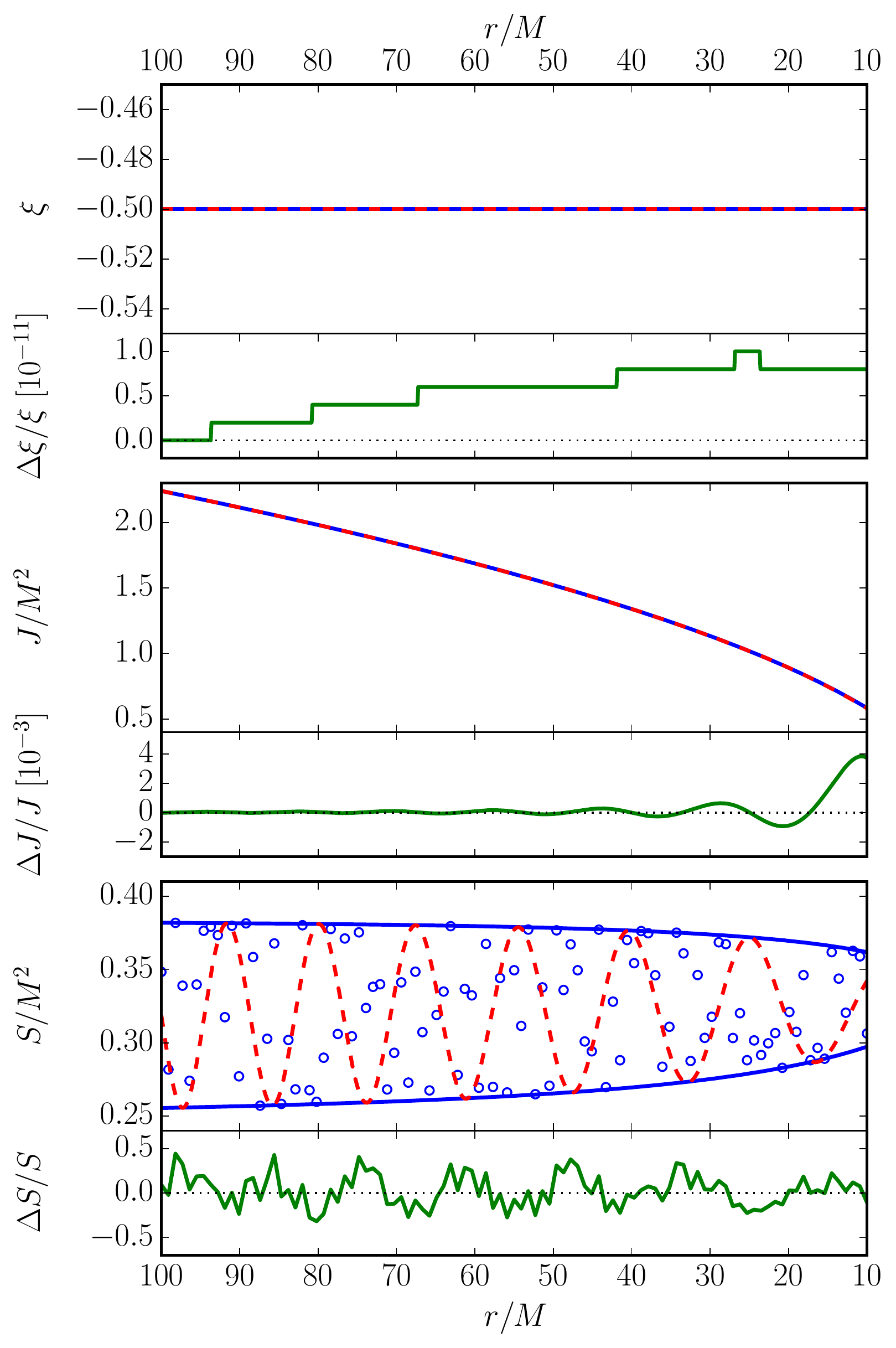}
\caption{Resulting plot obtained with the test function \fun{test.compare\_evolutions}, described in Sec.~\ref{compevol}. The source code is reported in Fig.~\ref{compevolcode}. We choose a BH  binary with $q=0.8$, $\chi_1=0.6$, $\chi_2=1$, $J=2.24 M^2$ and $\xi=-0.5$ at $r_i=100M$ and we compare its PN inspiral till $r_f=10M$ using  precession-averaged and orbit-averaged integrations. The evolutions of $\xi$ (top), $J$ (middle) and $S$ (bottom) are shown in the larger subpanels. Results for $J$ and $\xi$ show  excellent agreement between precession-averaged (solid blue) and orbit-averaged (dashed red) integrations. Precession-averaged integrations do not track the evolution of the total spin magnitude $S$, but  estimates (blue circles) can be obtained by sampling $S$ between $S_-$ and $S_+$ (blue solid lines); results are in statistical agreement with the orbit-averaged result (dashed red line). 
Smaller subpanels (solid green lines) show the relative difference between the two approaches.  This test is run typing  \comp{precession.test.compare\_evolutions()}.
}\label{compevolout}
\end{minipage}
\hfill
\begin{minipage}[b]{0.99\columnwidth}
\includegraphics[width=\columnwidth]{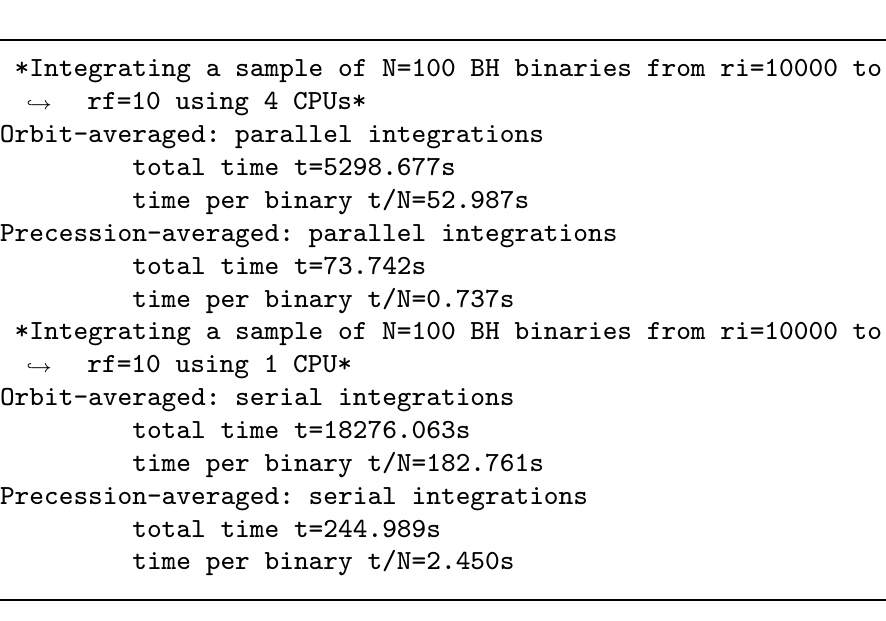}
\caption{Screen output of \fun{test.timing}, described in Sec.~\ref{timing}. The source code is reported in Fig.~\ref{timingcode}.
We time the performances of \fun{orbit\_angles} and \fun{evolve\_angles} using both parallel (first iteration) and serial (second iteration) computation.
Times reported here are obtained using a 2013 Intel i5-3470 3.20GHz 4 cores CPU. Output lines regarding the location of the stored data files are omitted for clarity. This test is run typing \comp{precession.test.timing()}. \label{timingout}}
\vspace{10cm}
\end{minipage}
\end{figure*}

Our last example, \fun{test.timing},  compares the computational efficiency of the PN integrators implemented in \precession. The test is executed with
\begin{verbatim}
    precession.test.timing()
\end{verbatim}
The source code and the screen output are reported in Figs.~\ref{timingcode} and \ref{timingout}, respectively.

 We compute the CPU time needed to evolve a sample of $N=100$ BH binaries from $r_i=10^4M$ to $r_f=10M$ using orbit-averaged and precession-averaged integrations. In particular, we time the orbit-averaged integrator wrapped inside \fun{orbit\_angles} (cf.~Sec.~\ref{orbav}) against the precession-averaged evolution implemented in \fun{evolve\_angles} (cf.~Sec.~\ref{precav}). The latter includes both the numerical integration of Eq.~(\ref{dJdr}) and a final resampling of the magnitude $S$.  To better illustrate the parallel implementation of the integrators, we perform the same computation twice: in the first iteration, integrations are performed in parallel on all the available cores (default); in the second iteration, we enforce a strictly serial execution. On average, a single BH inspiral  takes  $\sim 3$ minutes ($\sim 3$ seconds) when evolved using orbit- (precession-) averaged integrations. The computational performances  obtained here are  in good agreement with \cite{2015PhRvD..92f4016G}, where the dependence of the CPU time on the initial separation $r_i$ is also studied (see their Fig.~9).

\section{Conclusions}
\label{concl}

We have presented  design and  usage of the numerical open-source code \precession. Our code provides various numerical tools to study the precessional dynamics of BH binaries, evolve BH binaries along their GW-driven inspirals and estimate the properties of the single BHs resulting from binary mergers.
\precession\ is distributed  in the form of a \python\ module to combine flexibility, ease-of-use and numerical efficiency.  The code can be straightforwardly installed from the \href{https://pypi.python.org/pypi/precession}{\python\ Package Index} 
through \comp{pip},
 and it is distributed under version control at \href{https://github.com/dgerosa/precession}{{github.com/dgerosa/precession}}. Extensive documentation is regularly maintained at \href{https://dgerosa.github.io/precession}{{dgerosa.github.io/precession}}. Further information is available at \href{http://davidegerosa.com/precession}{{davidegerosa.com/precession}}.

\precession\ is under active development and several features will be added in future versions. Possible extensions include (i) generalization to eccentric orbits, (ii) explicit treatment of single-spin and non-spinning binaries, (iii) reparametrization of the dynamics in the equal-mass limit \cite{VosmeraPrep}, (iv) implementation of the latest fitting formulas to numerical-relativity simulations,
(v) addition of higher-order PN corrections,
and (vi) inclusion of numerical tools to study the resonant configurations $\alpha=2\pi n$ \cite{ZhaoPrep}. 
On the computational side, \precession\ will be ported to \python\ 3, and its parallel computing features further refined. Additional computational speed-up could be achieved using static compilers such as \textsc{cython} \citep{cython}. Compatibility and/or integration with the LIGO Algorithm Library\footnote{LAL, \url{www.lsc-group.phys.uwm.edu/lal}.} software is also an important future development.

The numerical tools described in this paper facilitate the implementation of spinning BH binary inspirals in a variety of astrophysical studies, ranging from population synthesis models to galaxy merger trees. Moreover, \precession\ provides flexible tools to interpret GW observations and numerical-relativity simulations of BH binaries in light of multitimescale PN techniques. As merging BH binaries have entered the realm of observations, we hope that our numerical efforts --here made available to the scientific community-- will help in understanding these fascinating physical systems straddling the boundaries between fundamental physics and astronomy.\\

 \acknowledgements 
We are grateful to Ulrich Sperhake, Emanuele Berti, Richard O'Shaughnessy, Alberto Sesana, Daniele Trifir\'o, Antoine Klein, Tyson Littenberg, Jakub Vosmera, Xinyu Zhao, Will Farr, Enrico Barausse and Guillame Faye for several fruitful discussions. This work was inspired by \cite{2015ApJS..216...29B}. D.G. is supported by the UK STFC and the Isaac Newton Studentship of the University of Cambridge. Partial support is also acknowledged from the Royal Astronomical Society, Darwin College of the University of Cambridge, the Cambridge Philosophical Society, the H2020 ERC Consolidator Grant No. MaGRaTh--646597, the H2020-MSCA-RISE-2015 Grant No. StronGrHEP-690904, the STFC Consolidator Grant No. ST/L000636/1, the SDSC Comet and TACC Stampede clusters through NSF-XSEDE Award No.~PHY-090003, the Cambridge High Performance Computing Service Supercomputer Darwin using Strategic Research Infrastructure Funding from the HEFCE and the STFC, and DiRAC's Cosmos Shared Memory system through BIS Grant No.~ST/J005673/1 and STFC Grant
No.~ST/H008586/1, and No. ST/K00333X/1. M.K. is supported by Alfred P. Sloan Foundation Grant No. FG-2015-65299 and NSF Grant PHY-1607031. This work was made possible by the open-source programming language \python\ \citep{Python} and the related tools \textsc{numpy} \citep{Walt}, \textsc{scipy} \citep{Jones:2001aa}, \textsc{matplotlib} \citep{2007CSE.....9...90H}, \textsc{parmap} \citep{Oller} and  \textsc{pdoc} \citep{pdoc}. Version-control distribution through \textsc{git} and \textsc{github} is also acknowledged. 
\newpage{}
\bibliography{coderelease}

\end{document}